\documentclass[pdflatex,sn-mathphys-num]{sn-jnl}

\usepackage[english]{babel}
\usepackage[T1]{fontenc}
\usepackage[utf8]{inputenc}
\usepackage{amsfonts}
\usepackage{amsmath}
\usepackage[most]{tcolorbox}
\usepackage{bbding}
\usepackage{fontawesome5}
\usepackage{graphicx}
\usepackage{amssymb}
\usepackage{wasysym}
\usepackage{float}
\usepackage{tikz-cd}
\usepackage{tikz}
\usetikzlibrary{shapes.geometric,arrows.meta,positioning}
\usepackage{physics}
\usepackage{palatino}
\usepackage{adjustbox}
\usepackage{multirow}
\usepackage{enumerate}
\usepackage{array}
\usepackage{amsmath}
\usepackage{braket}
\usepackage{mathtools}
\usepackage{comment}
\usepackage{xcolor}
\usepackage{pifont}
\usepackage{oplotsymbl}
\usepackage{subfig}
\usepackage{booktabs}
\usepackage{multirow}
\usepackage{tabularx}
\usepackage{geometry}
\usepackage{enumitem}  

\newcommand{\ignore}[1]{}
\newcommand{\ba}{\begin{eqnarray}}
\newcommand{\ea}{\end{eqnarray}}

\tikzset{ameter style/.style={
  draw,
  inner sep=0.4pt,
  rectangle,
  font=\vphantom{A},
  minimum width=0.9em,
  minimum height=0.9em,
  line width=0.5pt,
  path picture={
    \draw[black] ([shift={(.06,.12)}]path picture bounding box.south west) to[bend left=50] ([shift={(-.06,.12)}]path picture bounding box.south east);
    \draw[black,-{Latex[scale=0.45]}] ([shift={(0,.06)}]path picture bounding box.south) -- ([shift={(+.09,-.03)}]path picture bounding box.north);
  }
}}

\newcommand{\Ameter}{%
  \tikz[baseline=(ameter.base)]{%
    \useasboundingbox (-0.3cm, -0.3cm) rectangle (0.3cm, 0.3cm);
    \node[ameter style] (ameter) at (0,-0.05cm) {};%
  }%
}

\definecolor{myfuchsia}{RGB}{224,92,148}
\definecolor{myred}{RGB}{251, 77, 85}
\definecolor{myyellow}{RGB}{255, 255, 0}
\definecolor{azzurrino}{RGB}{204,229,255}
\definecolor{azzurrinochiaro}{RGB}{230,245,255}
\definecolor{giallino}{RGB}{255,255,204}

\newcommand{\Triangle}{%
  \unskip%
  \begin{tikzpicture}[scale=1, baseline={(0,-0.02)}]%
    \filldraw[fill=myyellow, draw=black, line width=0.15pt, rounded corners=0.03cm]
      (0,0) -- (0.27,0) -- (0.135,0.2338) -- cycle;%
  \end{tikzpicture}%
  \ignorespaces%
}

\newcommand{\ITriangle}{%
  \unskip%
  \begin{tikzpicture}[scale=1, baseline={(0,-0.02)}]%
    \filldraw[fill=myyellow, draw=black, line width=0.15pt, rounded corners=0.03cm]
      (0,0.2338) -- (0.27,0.2338) -- (0.135,0) -- cycle;%
  \end{tikzpicture}%
  \ignorespaces%
}

\newcommand{\MTriangle}{%
  \unskip%
  \begin{tikzpicture}[scale=1, baseline={(0,-0.02)}]%
    \filldraw[fill=myyellow, draw=black, line width=0.15pt, rounded corners=0.03cm]
      (0,0) -- (0.27,0) -- (0.135,0.2338) -- cycle;%
    \draw[line width=0.1pt]
      (0.108,0.117) -- (0.162,0.117);%
  \end{tikzpicture}%
  \ignorespaces%
}

\newcommand{\IMTriangle}{%
  \unskip%
  \begin{tikzpicture}[scale=1, baseline={(0,-0.02)}]%
    \filldraw[fill=myyellow, draw=black, line width=0.15pt, rounded corners=0.03cm]
      (0,0.2338) -- (0.27,0.2338) -- (0.135,0) -- cycle;%
    \draw[line width=0.1pt]
      (0.108,0.117) -- (0.162,0.117);%
  \end{tikzpicture}%
  \ignorespaces%
}

\newcommand{\UTileCircle}{
\begin{tikzpicture}[scale=1, baseline={([yshift=-0.08cm]current bounding box.center)}]
    \fill[fill=myfuchsia] (0,0) circle[radius=0.17];
    \fill[fill=white] (0,0) circle[radius=0.11];
    \draw[line width=0.2pt] (0,0) circle[radius=0.11];
    
\end{tikzpicture}
}

\newcommand{\UTileCIRCLE}{
\begin{tikzpicture}[scale=1, baseline={([yshift=-0.08cm]current bounding box.center)}]
    \fill[fill=myfuchsia] (0,0) circle[radius=0.17];
    \fill[fill=black] (0,0) circle[radius=0.11];
\end{tikzpicture}
}

\newcommand{\UTileLcircle}{
\begin{tikzpicture}[scale=1, baseline={([yshift=-0.08cm]current bounding box.center)}]
    \fill[fill=myfuchsia] (0,0) circle[radius=0.17];

    \begin{scope}
        \clip (0,0) circle[radius=0.11];
        \fill[black] (-0.11,-0.11) rectangle (0,0.11);
        \fill[white] (0,-0.11) rectangle (0.11,0.11);
    \end{scope}

    \draw[line width=0.2pt] (0,0) circle[radius=0.11];
\end{tikzpicture}
}
\newcommand{\UTileRcircle}{
\begin{tikzpicture}[scale=1, baseline={([yshift=-0.08cm]current bounding box.center)}]
    \fill[fill=myfuchsia] (0,0) circle[radius=0.17];

    \begin{scope}
        \clip (0,0) circle[radius=0.11];
        \fill[white] (-0.11,-0.11) rectangle (0,0.11);
        \fill[black] (0,-0.11) rectangle (0.11,0.11);
    \end{scope}

    \draw[line width=0.2pt] (0,0) circle[radius=0.11];
\end{tikzpicture}
}

\newcommand{\HTileTriangle}{%
  \unskip%
  \begin{tikzpicture}[scale=1, baseline={([yshift=-0.10cm]current bounding box.center)}]%
    \fill[fill=myred] (0,0) circle[radius=0.17];%
    \begin{scope}[shift={(-0.135,-0.078)}]%
        \begin{scope}[xshift=0.010cm, scale=0.93]%
            \filldraw[fill=myyellow, draw=black, line width=0.15pt, rounded corners=0.03cm]
                (0,0) -- (0.27,0) -- (0.135,0.2338) -- cycle;%
        \end{scope}%
    \end{scope}%
  \end{tikzpicture}%
  \ignorespaces%
}

\newcommand{\UTileTriangle}{
\begin{tikzpicture}[scale=1, baseline={([yshift=-0.10cm]current bounding box.center)}]
    \fill[fill=myfuchsia] (0,0) circle[radius=0.17];

    \begin{scope}[shift={(-0.135,-0.078)}]
        \begin{scope}[xshift=0.010cm, scale=0.93]
            \filldraw[fill=myyellow, draw=black, line width=0.15pt, rounded corners=0.03cm]
                (0,0) -- (0.27,0) -- (0.135,0.2338) -- cycle;
        \end{scope}
    \end{scope}
\end{tikzpicture}
}

\newcommand{\MixedOne}{%
  \tikz[baseline=-0.0ex, scale=0.3]{
    \draw[thin, dashed, dash pattern=on 1pt off 1pt] (0,0) rectangle (1,1);
    \filldraw[fill=white, draw=black] (0.3,0.7) circle[radius=0.1];
    \filldraw[fill=black, draw=black] (0.7,0.3) circle[radius=0.1];
  }%
}

\newcommand{\MixedTwo}{%
  \tikz[baseline=-0.0ex, scale=0.3]{
    \draw[thin, dashed, dash pattern=on 1pt off 1pt] (0,0) rectangle (1,1);
    \filldraw[fill=white, draw=black] (0.3,0.7) circle[radius=0.1];
    \filldraw[fill=white, draw=black] (0.7,0.2) arc[start angle=270, end angle=450, radius=0.1];
    \filldraw[fill=black, draw=black] (0.7,0.4) arc[start angle=90, end angle=270, radius=0.1];
  }%
}

\newcommand{\MixedThree}{%
  \tikz[baseline=-0.0ex, scale=0.3]{
    \draw[thin, dashed, dash pattern=on 1pt off 1pt] (0,0) rectangle (1,1);
    \filldraw[fill=white, draw=black] (0.3,0.6) arc[start angle=270, end angle=450, radius=0.1];
    \filldraw[fill=black, draw=black] (0.3,0.8) arc[start angle=90, end angle=270, radius=0.1];
    \filldraw[fill=black, draw=black] (0.7,0.2) arc[start angle=270, end angle=450, radius=0.1];
    \filldraw[fill=white, draw=black] (0.7,0.4) arc[start angle=90, end angle=270, radius=0.1];
  }%
}

\newcommand{\MixedFour}{%
  \tikz[baseline=-0.0ex, scale=0.3]{
    \draw[thin, dashed, dash pattern=on 1pt off 1pt] (0,0) rectangle (1,1);
    \filldraw[fill=white, draw=black] (0.3,0.7) circle[radius=0.1]; 
    \filldraw[fill=black, draw=black] (0.7,0.7) circle[radius=0.1]; 
    \filldraw[fill=black, draw=black] (0.3,0.3) circle[radius=0.1]; 
    \filldraw[fill=black, draw=black] (0.7,0.3) circle[radius=0.1]; 
  }%
}

\newcommand{\XMixedFour}{%
  \tikz[baseline=-0.0ex, scale=0.3]{
    \draw[thin, dashed, dash pattern=on 1pt off 1pt] (0,0) rectangle (1,1);
    \filldraw[fill=white, draw=black] (0.3,0.7) circle[radius=0.1]; 
    \filldraw[fill=white, draw=black] (0.7,0.7) circle[radius=0.1]; 
    \filldraw[fill=white, draw=black] (0.3,0.3) circle[radius=0.1]; 
    \filldraw[fill=black, draw=black] (0.7,0.3) circle[radius=0.1]; 
  }%
}
\newtcolorbox{nota}[1][]{enhanced,
    before skip=2mm,after skip=3mm,
    boxrule=0.4pt,
    left=15mm, 
    right=2mm,top=1mm,bottom=1mm,
    colback=yellow!50,
    colframe=yellow!20!black,
    sharp corners,rounded corners=southeast,arc is angular,arc=3mm,
    underlay={%
        \path[fill=tcbcolback!80!black] ([yshift=3mm]interior.south east)--++(-0.4,-0.1)--++(0.1,-0.2);
        \path[draw=tcbcolframe,shorten <=-0.05mm,shorten >=-0.05mm] ([yshift=3mm]interior.south east)--++(-0.4,-0.1)--++(0.1,-0.2);
        \path[fill=yellow!50!black,draw=none] (interior.south west) rectangle 
            node[white,rotate=0]{Nota} 
            ([xshift=14mm]interior.north west); 
    },
    drop fuzzy shadow,#1}
    \newtcolorbox{qm}[1][]{enhanced,
    before skip=2mm,after skip=3mm,
    boxrule=0.4pt,
    left=15mm, 
    right=2mm,top=1mm,bottom=1mm,
    colback=yellow!50,
    colframe=yellow!20!black,
    sharp corners,rounded corners=southeast,arc is angular,arc=3mm,
    underlay={%
        \path[fill=tcbcolback!80!black] ([yshift=3mm]interior.south east)--++(-0.4,-0.1)--++(0.1,-0.2);
        \path[draw=tcbcolframe,shorten <=-0.05mm,shorten >=-0.05mm] ([yshift=3mm]interior.south east)--++(-0.4,-0.1)--++(0.1,-0.2);
        \path[fill=yellow!50!black,draw=none] (interior.south west) rectangle 
            node[white,rotate=0]{MQ} 
            ([xshift=14mm]interior.north west); 
    },
    drop fuzzy shadow,#1}
    
\newtcolorbox{remark}[1][]{enhanced,
    before skip=2mm,after skip=3mm,
    boxrule=0.4pt,
    left=15mm, 
    right=2mm,top=1mm,bottom=1mm,
    colback=yellow!50,
    colframe=yellow!20!black,
    sharp corners,rounded corners=southeast,arc is angular,arc=3mm,
    underlay={%
        \path[fill=tcbcolback!80!black] ([yshift=3mm]interior.south east)--++(-0.4,-0.1)--++(0.1,-0.2);
        \path[draw=tcbcolframe,shorten <=-0.05mm,shorten >=-0.05mm] ([yshift=3mm]interior.south east)--++(-0.4,-0.1)--++(0.1,-0.2);
        \path[fill=red!50!black,draw=none] (interior.south west) rectangle 
            node[white,rotate=0]{!} 
            ([xshift=14mm]interior.north west); 
    },
    drop fuzzy shadow,#1}

         \newtcolorbox{mat}[1][]{enhanced,
    before skip=2mm,after skip=3mm,
    boxrule=0.4pt,
    left=15mm, 
    right=2mm,top=1mm,bottom=1mm,
    colback=blue!10!white,
    colframe=blue!20!black,
    sharp corners,rounded corners=southeast,arc is angular,arc=3mm,
    underlay={%
        \path[fill=tcbcolback!80!black] ([yshift=3mm]interior.south east)--++(-0.4,-0.1)--++(0.1,-0.2);
        \path[draw=tcbcolframe,shorten <=-0.05mm,shorten >=-0.05mm] ([yshift=3mm]interior.south east)--++(-0.4,-0.1)--++(0.1,-0.2);
        \path[fill=blue!40!black!70!white,draw=none] (interior.south west) rectangle 
            node[white,rotate=0]{\S} 
            ([xshift=14mm]interior.north west); 
    },
    drop fuzzy shadow,#1}

\newtcolorbox{exercise}[1]{%
    colback=teal!10,
    coltitle=black,
    colframe=teal!30,
    fonttitle=\bfseries,
    title=#1, 
    boxrule=0pt,
    enhanced
    }

\tikzset{
    dtr/.style={
        draw,
        shape border rotate=180,
        regular polygon,
        regular polygon sides=3,
        node distance=2cm,
        minimum height=4em
    }
}

\tikzset{
    utr/.style={
        draw,
        shape border rotate=0,
        regular polygon,
        regular polygon sides=3,
        minimum size=0.8cm
    }
}

\newlist{tabenum}{enumerate}{1}
\setlist[tabenum]{label=\alph*), nosep, leftmargin=*, topsep=0pt, partopsep=0pt}

\begin{document}

\title[Article Title]{QTris: a pedagogical board game to teach Quantum Mechanics}

\author*[1,2]{\fnm{Alessandro} \sur{Amabile}}\email{alessandro.amabile@unina.it}
\author*[3,4]{\fnm{Maria} \sur{Bondani}}\email{maria.bondani@cnr.it}
\author[1]{\fnm{Immacolata} \sur{De Simone}}\email{i.desimone21@studenti.unisa.it}
\author[1,2]{\fnm{Michela} \sur{Nazzaro}}\email{michela.nazzaro@unina.it}
\author[1,2]{\fnm{Michele} \sur{Viscardi}}\email{michele.viscardi@unina.it}
\author[1,2]{\fnm{Alioscia} \sur{Hamma}}\email{alioscia.hamma@unina.it}

\affil[1]{
\orgdiv{Dipartimento di Fisica}, 
\orgname{Universit\`a degli Studi di Napoli ``Federico II''},
 \postcode{80125}, \city{Napoli},
 \country{Italy}}
\affil[2]{
\orgdiv{INFN Sezione di Napoli}, 
 \postcode{80125}, \city{Napoli}.
\country{Italy}}
\affil[3]{
\orgdiv{Istituto di fotonica e nanotecnologie}, 
\orgname{Consiglio Nazionale delle Ricerche - CNR-IFN}, 
 \postcode{22100}, \city{Como},
\country{Italy}}
\affil[4]{
\orgdiv{Como Lake Institute of Photonics}, 
\orgname{Università degli Studi dell’Insubria}, 
 \postcode{22100}, \city{Como}.
\country{Italy}}

\abstract{In this paper we introduce the new version of QTris, a board game designed to teach and learn Quantum Mechanics within the framework of Quantum Information and Computation. The key idea behind the game is that every game sequence simulates a process on a system of qubits. Thus, QTris can be effectively integrated as a pedagogical tool to teach Quantum Mechanics at high-school level following a two-state approach. After arguing in support of this latter approach, we describe QTris' basic rules and some of its possible extensions, emphasizing how the game mechanics puts in clear light key quantum concepts such as incompatibility, probabilistic measurement and unitary transformation. Moreover, we report on the results of a QTris-based educational activity which involved about 150 high-school students and provided encouraging preliminary indications that QTris can be a useful pedagogical platform to promote an immediate understanding of some key concepts of Quantum Mechanics.}
\maketitle

\section{Introduction}
The teaching of Quantum Mechanics (QM) in high school has been a subject of increasing interest in recent years. Traditionally reserved for university-level physics courses, QM started to be considered an essential component of pre-university education due to its foundational role in modern science and technology \cite{Krijt2017, Merzel2024a}. This shift reflects the growing importance of quantum technologies, such as quantum computing and cryptography \cite{flagship}. However, incorporating QM into high school curricula presents unique challenges, including determining which aspects should be introduced \cite{Stadermann2019}, the pedagogical approaches best suited for this purpose \cite{Verbraeken2024}, and the effectiveness of alternative teaching methods such as educational games \cite{Chiofalo2022}.

In this paper we introduce the latest version of QTris, an educational board game designed by one of the authors to teach and learn QM within the framework of Quantum Information Theory \cite{Watrous, NielsenChuang}. Previous versions of the game are described in References \cite{Bondani2024} and \cite{Bondani2026}. The new version of the game will be available soon. However, the fundamental rules are unchanged from the previous version, which can be downloaded for free \cite{qtris}. The focus of this paper is on why and how QTris can be used as a pedagogical tool to introduce the elements of QM at high-school level. The paper is structured as follows.

In Section 2 we make some preliminary comments on the problem of teaching QM at high-school level. Here we highlight some critical aspects of the traditional \emph{historical} approach to QM and argue in favor of the so-called \emph{two-state} or \emph{qubit-first} approach, which also grounds QTris' game design.

In Section 3 we outline in greater detail the pedagogical reconstruction of QM behind QTris' game design and summarize the key concepts of QM that, we argue, can be conveyed at high-school level with full clarity and rigour. 

In Section 4 we describe QTris' basic rules. In short, the key idea is that the rules of the game reproduce the consequences of the postulates of QM when applied to a system of  qubits. Thus, every QTris' game sequence effectively simulates a real quantum experiment or calculation performed on a system of qubits.

In Section 5 we describe the structural analogy between QTris and QM. In particular, we show how one can design game problems in QTris that are actually typical QM problems in disguise. 

In Section 6 we describe QTris' advanced rules and extensions. Here we also discuss where, how and why in some points the correspondence between QTris' game elements and QM mathematical framework partially breaks down.

In Section 7 we discuss some preliminary results about the effectiveness of QTris in conveying some key concepts of QM in a stand-alone educational activity. 

In Section 8 we summarize the contents of the paper and point to some future developments.
\section{On Teaching QM at high-school level}
\subsection{The historical approach}
The traditional way to introduce the concepts and ideas of QM at pre-university level is the so-called \emph{historical approach}. According to this approach, which is adopted by many European national standards for secondary education, the natural entry point to QM is its historical birth in the field of atomic physics and electromagnetism at the beginning of the XX century. The Italian National Guidelines for High Schools, for example, recommend for the last year of the scientific curriculum an account of the "crisis" of classical physics (e.g., the problem of black-body radiation and the photoelectric effect) and a description of other important ideas from the period 1900-1925, including Bohr atomic model, de Broglie's wave-particle relations and Heisenberg's uncertainty principle for positions and momenta \cite{Giliberti2024}. Similar topics are included in the curricula of other European countries \cite{Stadermann2019}. 

Although these topics provide essential background knowledge, they also introduce formidable challenges at different levels. From the physical point of view, it is evident that, in order to understand the first problems of QM and their solutions, it is necessary to understand \emph{why} classical theories could not account for those phenomena. This, in turn, requires a deep understanding of many concepts and methods from classical physics, such as Maxwell’s equations and statistical mechanics, which usually cannot be treated in sufficient detail at high-school level. From the mathematical point of view, both late-classical and early-quantum problems are formulated in terms of advanced mathematical formalisms, particularly functional analysis and partial differential equations, which are clearly beyond the reach of high school students \cite{Singh2008}. As a result, lacking the necessary mathematical background, in practice many teachers must limit themselves to qualitative statements about the "quantum nature" of all these new phenomena discovered at the "atomic level" and resort to all sort of metaphors and semi-classical analogies in a desperate attempt to clarify the subject. Sadly, such attempts are in most cases doomed to fail, or, even worst, to reinforce some of the most persistent misconceptions about QM \cite{Fischler}. Expressions like ``a particle can be in two places at the same time" or ``the cat is at the same time alive and dead", which are common jargon also among professional physicists, are evidently \emph{meaningless} (and, in some cases, plainly \emph{false}) if one does not put an extra-care in explaining their appropriate context and use. Moreover, this requires a familiarity with the mathematics of QM which is not available in high-school contexts. In this regard, we remark in passing that most teachers do not have by themselves a sufficient knowledge in QM to address the subtleties of the early ``quantum" problems. In Italy, for example, most physics teachers in secondary school have a background in pure mathematics, where QM is often included only as an optional class. 

Curiously, the historical approach can be criticized also from the very same \emph{historiographical} point of view. In fact, modern scholarship in history of QM agrees in labeling \emph{Quantum Mechanics} the axiomatic framework first developed by John Von Neumann between 1927 and 1932 and regarded still today as the ``standard" formulation of QM. In a trilogy of papers \cite{DuncanTrilogy} and a book \cite{VonNeumann} it was Von Neumann who clarified and solved many of the problems that still obscured the foundations of the new theory, namely the equivalence of matrix and wave mechanics under the general structure of Hilbert spaces and the distinction between pure and mixed states, essential to clarify the role of probability in the new theory. Duncan and James \cite{DuncanJanssen1, DuncanJanssen2} used the metaphor of scaffolding and arch to describe the construction of this theoretical edifice, which took more than 20 years and passed through various stages sometimes grouped together under the name of \emph{Old Quantum Theory} (OQT). Thus, according to this view, \emph{the historical approach to QM is not QM at all}, since it focuses mostly on concepts and models developed in the period from 1900-1925 and stops at the very threshold of what we call QM today. Indeed, it is remarkable that the usual \emph{postulates} of QM and their key \emph{consequences} (the existence of superposition, entanglement, unitary evolution etc.) are not even \emph{mentioned} in Italian and other European guidelines for secondary education. So, in short, when it comes to QM our curricula misses their target and silently turn from  \emph{physics} to \emph{history} of physics, i.e. to the exposition of models and ideas which have been partly incorporated and partly superseded by the mature framework of QM.

This ``historical turn" is very problematic from the educational point of view. In fact, it is generally acknowledged, if not implied, that the aim of high school physics is to learn \emph{theories} and \emph{how to use} them to describe, explain, and predict the occurrence of real natural phenomena. To do so, it is customary to start from simple problems and gradually move to more complex ones. Take, for example, Newtonian dynamics. To familiarize with how the theory describes the motion of bodies, one usually starts with simple and idealized systems such as frictionless inclined planes, or simple pendula oscillating in vacuum; then, in a more or less gradual fashion, one arrives at the ``big" problems, such as those involving universal gravitation, which are also those whose solution contributed to the historical establishment of Newton's theory \cite{OxfordPrincipia}.  As well known, it was in the domain of astronomy that Newton's theoretical framework showed its superiority over its rivals, namely the cartesian approaches prevailing at the time of the first edition of the Principia \cite{Guicciardini, Nauenberg}; and yet, today, no one dares to \emph{begin} Newtonian mechanics using astronomical problems. Thus, one may ask: why with QM it should be different? The above described "historical turn"  is sometimes justified by the argument that "proper" QM is "too hard" from both the physical and mathematical point of view, so that a qualitative but historically-oriented treatment is the best one can achieve.  Thus, according to this view, QM would be "harder" than, say, Newtonian mechanics, Maxwell's electromagnetism or any other topic usually treated in some detail in high-school. Is this view really tenable? In asking such question, it is helpful to distinguish between the \emph{rules} of a physical theory, i.e., its \emph{postulates} or \emph{axioms}, from their \emph{empirical consequences} when they are applied to real problems. In the case of QM, as well known, the \emph{consequences} of the postulates are very counter-intuitive; yet, the \emph{rules} of the theory can be expressed quite simply. On the other hand, the \emph{predictions} of Classical Mechanics (CM) evidently make some contact with our experience of moving bodies and may even appear "intuitive", but Newton's laws of motion - the basic \emph{postulates} of the theory - are far from being so. In fact, as well known, the greatest power of Newton's theory comes from ideas that are deeply counter-intuitive, like universal action at a distance (!) and the principle of "action and reaction". The very same notion of \emph{force} in the Newtonian system is indeed a very \emph{abstract} conception, which gave rise to endless debates both historically and pedagogically \cite{Hestenes, DeGandt}. Still, over time we learned how to speak about ``forces" in a clear and much less cumbersome way than Newton did in the \emph{Principia}. This was possible thanks to (1) a progressive translation of Newton's theoretical framework in a simpler and more powerful mathematical language, namely vector calculus, and (2) by a pedagogical reconstruction of Newton's framework that reflected a deeper understanding of his theory. Thus, over more than three centuries the core ideas of Newtonian physics have been polished, clarified and made accessible to wider and wider audiences, till they became an established part of our scientific culture. We strongly believe that the same sort of pedagogical reconstruction is a necessary preliminary step for any attempt to teach QM at high-school level in a meaningful way, and, more generally, to make quantum concepts accessible to a non-specialist audience. We think this is feasible, and contrary to often quoted maxims about the ``impossibility to understand QM", our claim is that the \emph{rules} of QM are simple enough to be fully understandable by people with a high-school level mathematical background.

\subsection{The qubit-approach}
In Europe, large collaborative efforts within the QTEdu Project \cite{QTEdu, competence}  aimed at defining and implementing effective strategies for QM education and foster scientific literacy to prepare students for emerging quantum technologies.
To address the challenges posed by the traditional historical approach, alternative routes have been proposed in which students explore the fundamental principles of QM using a more accessible mathematical framework \cite{Michelini2000, McIntyre2022}. In particular, research has highlighted the effectiveness of the so-called \emph{two-state} or \emph{qubit-first} approach, which takes a \emph{qubit} as the paradigmatic example of quantum system. Real examples of qubits are polarized photons, trapped ions, atoms in a Stern-Gerlach apparatus and many more.  Research indicates that these approaches may introduce QM concepts in an intuitive and accessible manner, avoiding complex mathematical formalisms \cite{Merzel2024b} while still effectively illustrating key ideas such as state superposition, measurement, and entanglement \cite{Toth, Albert, Bitzenbauer2024}.

Following this trend, we strongly believe that the qubit approach provides the appropriate framework to perform the pedagogical reconstruction that is required to introduce modern QM at high-school level, and, at the same time, to bypass the inherent difficulties of the historical approach. In our view, the reasons may be summarized as follows: 
\begin{enumerate}
    \item Focusing the attention on finite-dimensional systems allows to bypass the need for functional analysis and other advanced mathematical techniques. It is true that the problems addressed in the early history of QM involved continuous variables as position and momentum, but today we know that, for all practical purposes, there is \emph{no loss of generality} in restricting to finite-dimensional Hilbert spaces. From the pedagogical point of view, there is much to gain in simplicity and nothing to lose conceptually, since all the key elements of QM may be illustrated equally well. Thus, the only mathematics that is needed is \emph{linear algebra}, namely the algebra of linear operators on \emph{finite}, \emph{complex} vector spaces. This machinery, together with some basic elements of probability theory, is all one needs to construct with full rigour the theoretical framework of QM, in its most modern and powerful form.
     \item A qubit is the most elementary quantum system. Therefore,  from the pedagogical point of view, it is evident that a system of qubits - be they realized as polarized photons, trapped ions, atoms in a magnetic field, etc. - represent the ideal ground to familiarize with the consequences of the postulates, i.e. with the empirical content of QM. All the key elements of the theory - superposition, entanglement, unitary evolution, probabilistic measurements and so on - are fully exhibited by the behavior of qubits. Using an analogy, it may be said that qubits are for QM what inclined planes and pendula are for CM: elementary (and very idealized) systems that serve the purpose to learn \emph{how to use} some well-defined \emph{rules} to frame and solve some specific physical \emph{problems}, namely to describe and predict the results of real quantum experiments and computations. Exactly this purpose - and even more - can be achieved in QM using the qubit-first approach;
    \item The qubit-first approach makes contact with contemporary trends in theoretical and experimental quantum research. The rise of quantum information theory and the related application in quantum technologies caused a real paradigm shift in QM, turning physicists' attention from what the theory \emph{says} about the world to what the theory \emph{enables us to do} \cite{NielsenChuang}. Of course these two perspectives are not exclusive but illuminate each other; however, it is undeniable that the possibility of manipulating single quantum systems - something unconceivable only 50 years ago - really changed the way we perceive the "hard truth" of QM. No wonder, then, if such a \emph{theoretical} shift should turn out to be just as relevant from a \emph{pedagogical} point of view. 
    \end{enumerate}

\subsection{Game-based approaches to QM}
In this context, educational games have emerged as a powerful tool for teaching abstract and complex subjects such as QM \cite{Plass2015}. Games create an interactive environment in which students can experiment with quantum principles through visual and experiential learning rather than relying only on theoretical explanations \cite{Meyer1999, Piispanen2025}. Quantized versions of well-known games have appeared, starting from the simple coin-tossing discussed by Meyer \cite{Meyer1999}, and quantum approaches to fundamental problems in game theory, such as the Battle of the Sexes \cite{Nawaz2004}, have been reformulated to incorporate quantum superposition, altering the strategic equilibria \cite{Iqbal2002}. Similarly, the Monty Hall problem \cite{D'Ariano2002}, when recast in a quantum framework, illustrates the profound impact of quantum probabilities and entanglement on decision-making processes. 
 Online platforms such as qplaylearn \cite{qplaylearn} offer interactive games and educational materials to support the teaching of quantum mechanics through game-based methodologies. One of the most notable adaptations in this field is the quantum version of Tic-Tac-Toe — a universally known childhood game known under various names, including \emph{Tic-Tac-Toe} (U.S.), \emph{Xs and Os} (Canada), \emph{Noughts and Crosses} (U.K.), and \emph{Tris} (Italy). A quantum variant of \emph{Tris} was first proposed by Goff in 2002 \cite{goff2006} as an educational tool for high school students, offering an engaging and accessible way to introduce fundamental quantum principles. Since then, several versions of quantum \emph{Tic-Tac-Toe} have been developed, each with distinct rules and interpretations \cite{Weingartner2023}.

For our part, the idea to design a game to teach QM comes mainly from the inescapable role that \emph{abstraction} plays in QM. In fact, as well known, the rules of QM are quite clear as long as they are abstract, but get very complicated when one tries to connect them to something {\em concrete} as a real process occurring in an experiment or, even worse, to common sense. This request of concreteness is essential and should not be dismissed; however, it is often tainted with a sort of hostility towards abstraction \emph{per se}, that, in our view, is absolutely detrimental for physics teaching. On the contrary, we believe that one of the higher aims of physics education (and perhaps of education in general) is exactly to teach \emph{how to abstract}, i.e., to find a \emph{structure} lying behind the multifaceted variety of events occurring in our experience of the physical world. In theoretical physics the recognition of this ``abstract" structure is usually taken as indicating a deeper \emph{understanding} of the theory itself. So, it is with this view towards abstraction \emph{as} understanding that we believe young students can get the greatest payoff from a game-based approach to QM, since all games, by their very nature, are abstract. Take the example of chess: players know very well that the concrete appearance or interpretations of the pieces on the board is immaterial. The only things that matter are the \emph{rules of the game} played through chess pieces: how they can move, their configuration on the board and the possible strategies that, within the rules, may lead players to victory. Whether the piece called ``horse" really looks like a ``horse" is inessential: the horse shape is a make-believe, a fiction that players know very well play no role in their strategy. Kids, who take games seriously, know this very well; and \emph{this} is essential to understand physical theories in general and QM in particular as most physicists do in their everyday practice: as a \emph{set of principles} and \emph{rules} which, when applied correctly, describe, predict and, in some sense, \emph{explain} the behavior of physical systems under well-defined experimental conditions. As well known, in the case of QM, such rules have never failed us, making it perhaps the best-tested theory in the history of physics.
\section{Pedagogical Reconstruction of QM}
In this Section we outline the core concepts of QM that we think can be conveyed at high-school level with full clarity and rigour. As we will see in the following Section, QTris' rules are designed in such a way to put these key ideas in clear light.

The first thing to clarify in introducing QM is that, from the logical point of view, QM sits at a higher level than a theory for a new domain of physical phenomena. QM is something different: it is a whole new \emph{framework} for the description of physical phenomena, i.e., a new \emph{set of principles} for the construction of physical theories. Thus, from the logical point of view, introducing QM is quite different from, say, introducing electromagnetism on the context of classical physics. In that case, the general framework is given by the principles of CM and assumed to be known; then, the problem is to extend its application to extend the picture into a wider region of phenomena. On the other hand, the step from CM to QM is one of a very different kind, since it is the \emph{framework itself} that is \emph{expanded}, and not only its domain of physical application. The old or "classical" framework is included as a particular case of the new or "quantum" one, but it is the whole \emph{language} of the description that changes in non-trivial ways. 

To be as explicit and clear as possible about this shift in the \emph{grammar} of physical theories, we rely on the Ludwig-Kraus scheme for the description of quantum processes \cite{Kraus}. Within this scheme, any quantum experiment is analyzed in a tripartite way: \emph{preparation}, \emph{(unitary) transformation} and \emph{measurement}. 

A \emph{preparation} is an idealized experimental protocol whose output is  an \emph{ensemble} of physical systems with some well-defined physical properties. In all real experiments the predictions of QM are related to the properties of systems extracted by an ensemble of systems whose physical properties are encoded by a mathematical object called \emph{state}. Following common use, we will often speak at the singular and write "state of \emph{the} system", but it is always implied that the system in question is extracted from a well-prepared ensemble described by a certain quantum state. So, in short, in QM a \emph{state} is a mathematical object which encodes the physical properties of an ensemble of systems resulting from of a well-defined \emph{preparation}. When the preparation is such that all the systems in the ensemble are \emph{identical}, (i.e., they all have \emph{identical} physical properties), the associated state is called \emph{pure}; in all other cases the state is called \emph{mixed}. In real applications states are always mixed, since any non-idealized process of preparation is affected by experimental uncertainty.

Through interactions quantum systems may change their properties and, thus, their state, according to exact and deterministic rules. Such transformations are called \emph{unitary transformations}. As well known, the class of unitary transformations plays a crucial role in QM. For example, it is a fundamental postulate of the theory that the evolution of any isolated system is described by a unitary transformation.\footnote{As well known, perfect isolation is an idealized condition that can be realized in practice only in an approximate way. Isolation from the environment is indeed one of the key problems of all quantum technologies.} Moreover, if a state is pure, it can be transformed into any other pure state using only unitary transformations. 

A \emph{measurement} is the experimental equivalent of \emph{asking a question} to a system about one or more of its physical properties. Every question can be put into binary terms, and when measured the system is obliged (in practice \emph{forced}) to answer. In some cases, after measurement the system is destroyed; in other cases, the measurement changes the state and it is possible to perform other operations. One of the main objectives of QM as a physical theory is to predict the results of \emph{any} measurements that can possibly be performed on \emph{any} quantum system.

Given the general scheme outlined above, there are four key concepts of QM that we wish to convey. They are the following:

\paragraph{A. QM is a \emph{probabilistic} theory.} The only things the theory predicts for any measurement are the \emph{probabilities} associated to the possible outcomes. All probabilities of all possible measurements are encoded in the state of the system. A notable feature of QM is that its validity does not depend on the approach to probability one chooses to adopt. In many experimental situations it is customary to follow the \emph{frequentist} definition, so that probabilities are defined as a \emph{theoretical limit} that is indefinitely approached as one repeats the \emph{same} measurement many times on systems resulting from the same preparation. However, there are also situations in which a system is prepared and measured only once, so that a \emph{subjectivist} or \emph{Bayesian} approach is required to make sense of a probabilistic statement. Whatever the approach may be, \emph{certainty} is included in any probabilistic setting as the case corresponding to a maximal value ($0$ or $1$) of the probability. 

\paragraph{B. In QM probabilities can be modified in a \emph{deterministic} way.} This point must be emphasized, since it is often overlooked: in QM, the appearance of probabilities in the description of physical phenomena \emph{is not the whole story}. After all, if this was the case, QM would not be different from, say, classical stochastic mechanics, which also deals with probability distributions over ensembles of physical systems. Where is the "quantum revolution" then? Of course, it is true that in QM probability plays a central role, but the \emph{essential} feature of probabilities arising from quantum phenomena is that they are \emph{subject to deterministic laws}: whenever a system undergoes a unitary process, its state and all the measurement probabilities it encodes change in a completely deterministic way. From a more operational perspective, it is a consequence of QM that, if a quantum system is in a pure state, acting on it with suitable instruments we can modify \emph{at will} the probability distributions observed under measurement. This is the case, for example, of spins precessing in magnetic fields, of optical plates rotating the polarization of photons, and of logical gates in quantum processors, which are all different examples of physical interactions represented by unitary operations. In fact, this complete control over probabilities of physical events - which is impossible in classical technologies, where probabilities arise only as a way to cope with our ignorance of some property of the system - is what makes \emph{in principle} quantum technologies more powerful than classical ones. As we will see, one of the most valuable features of QTris is that it emphasizes the conceptual and operational importance of \emph{unitary operations}, since the whole game strategy is based on them.

\paragraph{C. In nature there exist \emph{incompatible} physical properties.} Two properties $A$ and $B$ are (maximally) \emph{incompatible} if, given the value of $A$, measuring $B$ may give any one of its possible values with equal probability (and \emph{viceversa}). Also \emph{partial} incompatibility is possible, being the case when, given the value of $A$, $B$ has a non-uniform probability distribution over its possible values (and \emph{viceversa}). Not only incompatible properties exist, but it turns out that for \emph{every} physical system there are couples of properties which exhibit this behavior under measurement. Typical examples are the corresponding components of position and momentum of a particle in space, the orthogonal components of the spin of an electron in a magnetic field, and the orthogonal components of the polarization of a photon. The existence of incompatibility, formalized as non-commutativity between quantum observables, is at the root of Heisenberg uncertainty relations and one of the essential points in which the ontology of QM departs from that of classical physics \cite{StanfordUncertainty}.\footnote{Notice that the widespread belief that Heisenberg uncertainty relations derive solely from the some "disturbance" induced by measurement is false.} Incompatibility is indeed what gives physical meaning of the often-quoted phrase that quantum probabilities are "essential": probabilities cannot be \emph{removed} from the theory, since to describe \emph{all possible properties} of physical systems one must include also incompatible ones. In this sense, one can say that QM is \emph{more general} than CM, since this latter does not admit the existence of incompatible properties. 

\paragraph{D. Incompatibility and composition lead to \emph{entanglement}.} In the case of composite and interacting systems, the existence of incompatible properties gives rise to \emph{entanglement}. Therefore, \emph{entanglement}-related phenomena are impossible in any classical theory which does not admit the existence of incompatible properties. Experimentally, entanglement manifests itself as a correlation between the effects observed by acting on the different subsystems. However, like before, in quantum \emph{entanglement} the appearance of correlations is not the whole story. After all, correlated results of measurements on different systems are quite natural if those systems have interacted in the past, and there is nothing inherently "quantum" in this behavior. So, what is the difference between classical correlations and quantum entanglement? Again, the essential point is that, \emph{if the state is pure, correlations arising from entanglement can be modified at will}. Thus, for example, entangled subsystems whose properties are perfectly correlated may become perfectly anticorrelated (and \emph{viceversa}). More generally, as with probability distributions, if the composite state is pure correlations can be modified in an infinite number of ways, opening a world of technological possibilities that are unavailable in classical technologies and that are still being explored. Indeed, in the last decades entanglement has been recognized as one of the most powerful resources for the implementation of quantum technologies. The huge amount of misleading narratives developed around the concept of entanglement makes very urgent to find ways of explaining entanglement-related phenomena that are clear, rigorous and not ambiguous. Just to make an example, the possibility of a "spooky action at a distance" based on entanglement has been excluded both theoretically \cite{Eberhard, Ghirardi, nonlocality} and experimentally \cite{Sciarrino}. Therefore, any exposition that leaves room for doubt about this point must be forcefully rejected as mystification. As we will see, QTris offers a simple and rigorous way to convey the conceptual and operational meaning of entanglement without indulging into misleading metaphors.

\section{QTris: the rules of the game}
In this Section we summarize the basic rules of QTris. As we will see in Section 6, QTris can be expanded in multiple ways, but here we will focus only on its fundamental structure. Moreover, in this Section we will omit any consideration of the connections between QTris and QM, which will be explained in detail in the following. However, any one who knows QM will not fail to grasp such connections. 

\begin{figure}
    \centering
\includegraphics[width=0.5\textwidth]{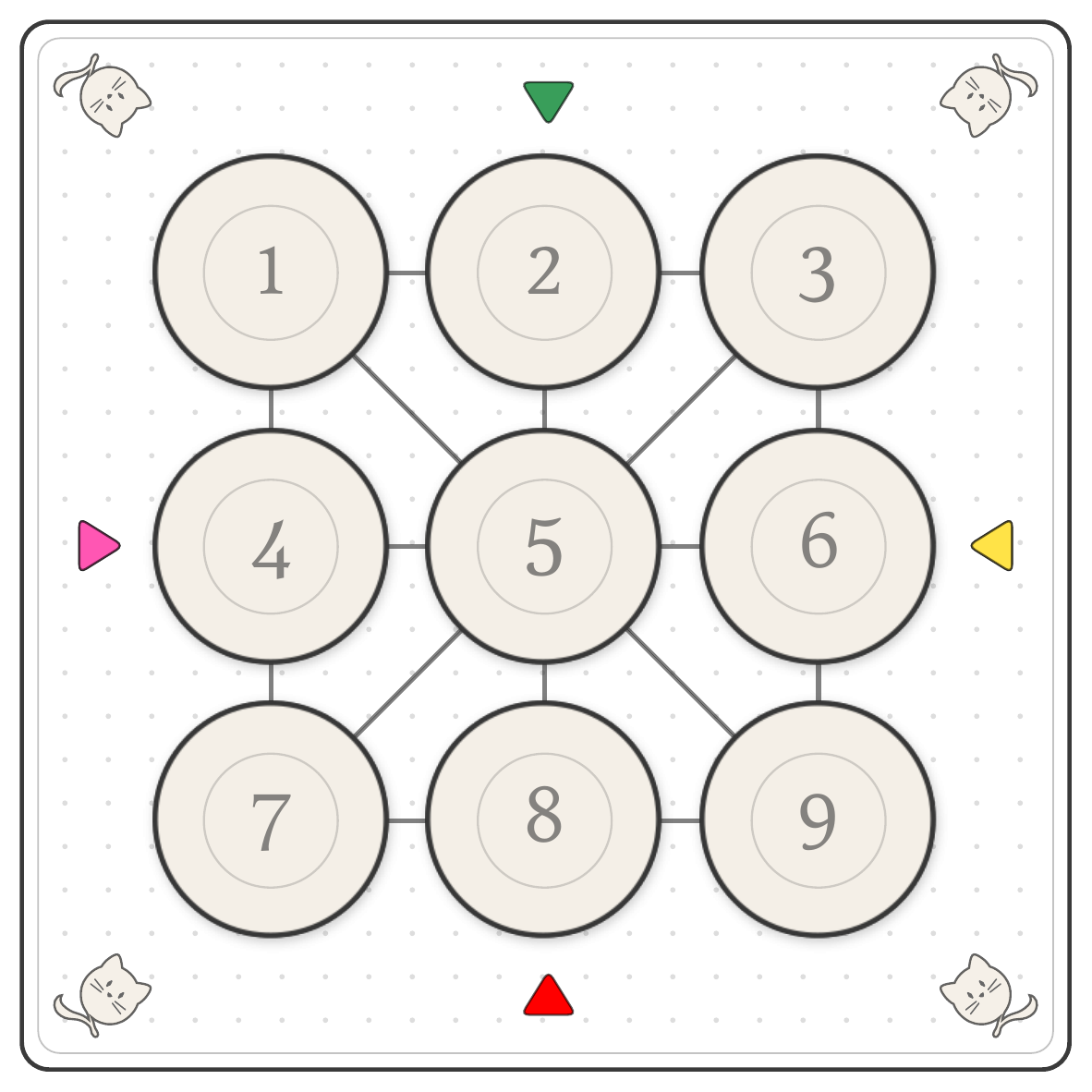}
    \caption{The 3x3 game board of QTris, the same as that of classical tic-tac-toe.}
        \label{fig:board}
\end{figure} 

QTris, abbreviation of \emph{Quantum Tris}, is a variant of the traditional \emph{Tic-Tac-Toe}, known in Italy as \emph{Tris}. There are two players, who play respectively for "white" and "black". Like in Tris, the board is a 3x3 grid (Fig.~\ref{fig:board}) and the aim of the players is to align three tiles of their color.

Every game is divided into three phases:
\begin{enumerate}
\item \emph{Preparation phase.} The grid is filled with tiles according to some protocol. Normally, tiles are placed at random on the board, but one can choose to start with any given configuration. When a given tile is placed on a cell, we say that the cell is in the corresponding \emph{state}. In the basic version of the game, there are four possible states, represented by four different tiles:
\begin{enumerate}
    \item \emph{white}, or $\Circle$; 
    \item \emph{black}, or $\CIRCLE$;
    \item \emph{left}, or $\LEFTcircle$;
    \item \emph{right}, or $\RIGHTcircle$.
\end{enumerate}
In the following, we will sometimes refer to $\{\Circle, \CIRCLE\}$ as \emph{color-tiles} and to $\{\LEFTcircle, \RIGHTcircle\}$ as \emph{orientation-tiles}. Moreover, in the preparation phase players receive 4 move cards, that will be used in the next phase.
\item \emph{Operations phase}. In this phase players use cards to modify the tiles on the board. In the basic version of QTris there are five cards:
\begin{itemize}
    \item Card $I$ or \emph{identity}. The effect of this card is trivial: it leaves every tile unchanged;
    \item Card $X$ or \emph{Color flip}. It turns \emph{white} into \emph{black} and viceversa. In symbols,
    \[
    \Circle \xleftrightarrow{X} \CIRCLE.
    \]
    On orientation-tiles, it acts like $I$;
    \item Card $Z$ or \emph{Orientation flip}. It turns \emph{left} into \emph{right} and viceversa. In symbols,
    \[
    \LEFTcircle \xleftrightarrow{Z} \RIGHTcircle.
    \]
    On color-tiles, it acts like $I$;
    \item Card $Y$ or \emph{total flip}. It changes both \emph{Color} and \emph{Orientation}, turning \emph{white} into \emph{black} (and viceversa) and \emph{left} into \emph{right} (and viceversa). In symbols,
    \begin{align*}
    \Circle &\xleftrightarrow{Y} \CIRCLE; \\
    \LEFTcircle &\xleftrightarrow{Y} \RIGHTcircle.
    \end{align*}
    In short, $Y$ acts like $X$ or $Z$ according to the tile it acts upon;
    \item Card $H$ or \emph{Hadamard}. It turns \emph{white} into \emph{left} (and viceversa) and \emph{black} into \emph{right}. In symbols:
    \begin{align*}
    \Circle \xleftrightarrow{H} \LEFTcircle; \\
    \CIRCLE \xleftrightarrow{H} \RIGHTcircle.
    \end{align*}
    \end{itemize}
The action of the cards $X, Y, Z$ and $H$ is represented by the square map in Fig.\ref{fig:algebra_1}. 
\begin{figure}
    \centering
    \includegraphics[width = 0.5\textwidth]{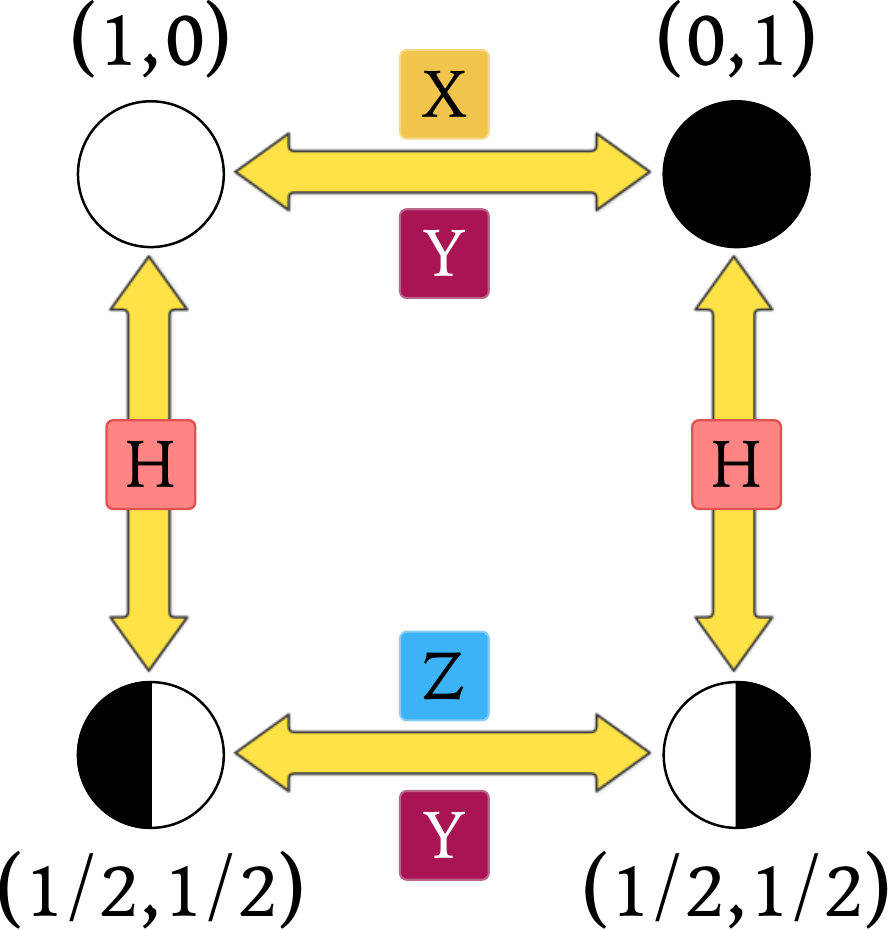}
    \caption{\textbf{Square map of operations $X, Y, Z, H$ and of measurement probabilities.} 
    The numbers in the parentheses $(x,y)$ next to the symbols $\Circle,\CIRCLE, \LEFTcircle, \RIGHTcircle$ represent the probabilities of obtaining $\Circle$ and $\CIRCLE$ respectively, after performing a measurement. Two symbols on and below an arrow mean that both operations achieve the same effect.}
    \label{fig:algebra_1}
\end{figure}
In the operation phase players take turns, and at every turn they (1) play 2 cards and (2) take 2 other cards from the deck, so that they always have 4 cards. The phase terminates after 5 turns. Notice that if during this phase three tiles of the same color are aligned no point is scored. As we will see shortly, players score point only at the end of the last phase of the game (see example below).

\item \emph{Measurement phase}. For each cell of the grid players roll a die to \emph{measure} its color, which can be either \emph{white} or \emph{black}. After measurement, the tile on the cell is changed according to the result, so that at the end of the measurement phase every tile is either black or white. Thus, players score one point for each row, column or diagonal of their color and the game is over. For each possible state the rule-book gives the probabilities for the two possible results, and the corresponding division of the set of possible outcomes of the die roll. The measurement probabilities of \emph{white} and \emph{black} for the four basic tiles are shown in parentheses in Fig.~\ref{fig:algebra_1}. For example, if the state is $\LEFTcircle$ and one uses a d100, the result of the color measurement will be $\Circle$ if the outcome lies between 1 and 50 (included), and $\CIRCLE$ if the outcome lies between 51 and 100 (included).\footnote{In practice players use two d10s numbered from $0$ to $9$, with the first die giving the \emph{units} and the second one giving the \emph{tens}. For example, if the first die gives $4$ and the second gives $9$, the number is $94$. The number $100$ corresponds to outcomes $0$ and $0$.} We will represent a measurement process by the symbol $\Ameter$. For example, if a color measurement on the tile $\LEFTcircle$ gives the result $\CIRCLE$ we will write 
\[
\LEFTcircle \xrightarrow{\Ameter} \CIRCLE.
\]
\end{enumerate}
In short, the aim of the players is to use the cards at their disposal to (1) maximize their probability to make a point and (2) minimize the probability of the opponent to do the same.

Below, it is presented a complete example of the game, with all its phases. 
\begin{enumerate}
    \item \emph{Preparation}. White, black, left and right tiles are placed on the numbered cells of the initially empty game board. As an example, the game board is prepared as follows.
 \begin{align*}
 \begin{pmatrix}
1 & 2 & 3 \\
4 & 5 & 6 \\
7 & 8 & 9
\end{pmatrix}
\;\xrightarrow\;
        {     \renewcommand{\arraystretch}{1}     \begin{pmatrix}
            \LEFTcircle & \Circle &\RIGHTcircle\\
       \CIRCLE &\CIRCLE &\LEFTcircle \\
       \LEFTcircle &\RIGHTcircle &\CIRCLE
        \end{pmatrix}     }
    \end{align*}
Black and White players pick 4 cards. White starts with the cards $X, Y, H, I$, while Black starts with the cards $X, Z, H, H$.
\item \emph{Operations}. Players take turns and use their cards to modify the configuration of tiles on the board. White takes the first turn.
\begin{itemize}
\item \textbf{Turn $1$:} White plays $H$ on cell $1$ and $X$ on cell $4$. They draw cards $I, H$. Black plays $X$ on cell $1$ and $Z$ on cell $7$. They draw cards $H, Y$.
    At the end of the turn, the game board looks like:
     \begin{align*}
        {     \renewcommand{\arraystretch}{1} 
        \begin{pmatrix}
            \LEFTcircle & \Circle &\RIGHTcircle\\
       \CIRCLE &\CIRCLE &\LEFTcircle \\
       \LEFTcircle &\RIGHTcircle &\CIRCLE
        \end{pmatrix}  \;\xrightarrow{W}\;  \begin{pmatrix}
            \Circle & \Circle &\RIGHTcircle\\
       \Circle &\CIRCLE &\LEFTcircle \\
       \LEFTcircle &\RIGHTcircle &\CIRCLE
        \end{pmatrix}  \;\xrightarrow{B}\;  
        \begin{pmatrix}
            \CIRCLE & \Circle &\RIGHTcircle\\
       \Circle &\CIRCLE &\LEFTcircle \\
       \RIGHTcircle &\RIGHTcircle &\CIRCLE
        \end{pmatrix}     }
    \end{align*}
    Notice that there are three black tiles along a diagonal, but this is irrelevant: points are given only at the end of the measurement phase.
  \item \textbf{Turn $2$:} White plays two times the $I$ card. They draw cards $X, H$. Black plays $H$ on cell $7$ and $H$ on cell $8$. They draw cards $I, Y$.
    At the end of the turn, the game board looks like:
     \begin{align*}
        {     \renewcommand{\arraystretch}{1} 
        \begin{pmatrix}
            \CIRCLE & \Circle &\RIGHTcircle\\
       \Circle &\CIRCLE &\LEFTcircle \\
       \RIGHTcircle &\RIGHTcircle &\CIRCLE
        \end{pmatrix}       \;\xrightarrow{W}\;  \begin{pmatrix}
            \CIRCLE & \Circle &\RIGHTcircle\\
       \Circle &\CIRCLE &\LEFTcircle \\
       \RIGHTcircle &\RIGHTcircle &\CIRCLE
        \end{pmatrix}     \;\xrightarrow{B}\;  
        \begin{pmatrix}
            \CIRCLE & \Circle &\RIGHTcircle\\
       \Circle &\CIRCLE &\LEFTcircle \\
       \CIRCLE &\CIRCLE &\CIRCLE
        \end{pmatrix}     }
    \end{align*}
    
  \item \textbf{Turn $3$:} White plays $H$ on cell $3$ and $X$ on cell $8$. They draw cards $Y, Z$. Black plays $I$ and then $Y$ on cell $4$. They draw cards $I, X$.
    At the end of the turn, the game board looks like:
     \begin{align*}
        {     \renewcommand{\arraystretch}{1} 
        \begin{pmatrix}
            \CIRCLE & \Circle &\RIGHTcircle\\
       \Circle &\CIRCLE &\LEFTcircle \\
       \CIRCLE &\CIRCLE &\CIRCLE
        \end{pmatrix}       \;\xrightarrow{W}\;  \begin{pmatrix}
            \CIRCLE & \Circle &\LEFTcircle\\
       \Circle &\CIRCLE &\LEFTcircle \\
       \CIRCLE &\Circle &\CIRCLE
        \end{pmatrix}     \;\xrightarrow{B}\;  
        \begin{pmatrix}
            \CIRCLE & \Circle &\LEFTcircle\\
       \CIRCLE &\CIRCLE &\LEFTcircle \\
       \CIRCLE &\Circle &\CIRCLE
        \end{pmatrix}     }
    \end{align*} 

   \item  \textbf{Turn $4$:} White plays $H$ on cell $3$ and $Y$ on cell $5$. They draw cards $I, X$. Black plays $X$ on cell $5$ and $Y$ on cell $6$. They draw cards $I, H$.
    At the end of the turn, the game board looks like:
     \begin{align*}
        {     \renewcommand{\arraystretch}{1} 
        \begin{pmatrix}
            \CIRCLE & \Circle &\LEFTcircle\\
       \CIRCLE &\CIRCLE &\LEFTcircle \\
       \CIRCLE &\Circle &\CIRCLE
        \end{pmatrix}       \;\xrightarrow{W}\;  \begin{pmatrix}
            \CIRCLE & \Circle &\Circle\\
       \CIRCLE &\Circle &\LEFTcircle \\
       \CIRCLE &\Circle &\CIRCLE
        \end{pmatrix}     \;\xrightarrow{B}\;  
        \begin{pmatrix}
            \CIRCLE & \Circle &\Circle\\
       \CIRCLE &\CIRCLE &\RIGHTcircle \\
       \CIRCLE &\Circle &\CIRCLE
        \end{pmatrix}     }
    \end{align*}

    \item \textbf{Turn $5$:} White plays $X$ on cell $4$ and $Y$ on cell $5$. Black plays $H$ on cell $3$ and $H$ on cell $5$. Cards are not drawn in the last turn.
    At the end of the turn, the game board looks like:
     \begin{align*}
    \renewcommand{\arraystretch}{1}
    \begin{pmatrix}
        \CIRCLE & \Circle & \Circle \\
        \CIRCLE & \CIRCLE & \RIGHTcircle \\
        \CIRCLE & \Circle & \CIRCLE
    \end{pmatrix}
    \;\xrightarrow{W}\;
    \begin{pmatrix}
        \CIRCLE & \Circle & \Circle \\
        \Circle & \Circle & \RIGHTcircle \\
        \CIRCLE & \Circle & \CIRCLE
    \end{pmatrix}
    \;\xrightarrow{B}\;
    \begin{pmatrix}
        \CIRCLE & \Circle & \RIGHTcircle \\
        \Circle & \LEFTcircle & \RIGHTcircle \\
        \CIRCLE & \Circle & \CIRCLE
    \end{pmatrix}
\end{align*}
    
\end{itemize}

\item  \emph{Measurement}. At the end of the last turn, the game board looks like: 
\begin{align*}
\begin{pmatrix}
        \CIRCLE & \Circle & \RIGHTcircle \\
        \Circle & \LEFTcircle & \RIGHTcircle \\
        \CIRCLE & \Circle & \CIRCLE
    \end{pmatrix}
\end{align*}
The measurement is performed only on the cells which are not occupied by color tiles. The d$100$ die is rolled, resulting in the following outcomes: 
\begin{itemize}
    \item On cell $3$, the result is $53$. The tile $\RIGHTcircle$ is replaced by $\CIRCLE$.
    \item On cell $5$, the result is $7$. The tile $\LEFTcircle$ is replaced by $\Circle$. 
    \item On cell $6$, the result is $89$. The tile $\RIGHTcircle$ is replaced by $\CIRCLE$.
\end{itemize}
After measurement, the game board looks like: 
\begin{align*}
\begin{pmatrix}
        \CIRCLE & \Circle & \RIGHTcircle \\
        \Circle & \LEFTcircle & \RIGHTcircle \\
        \CIRCLE & \Circle & \CIRCLE
    \end{pmatrix} \xrightarrow{\Ameter} 
\begin{pmatrix}
        \CIRCLE & \Circle & \CIRCLE \\
        \Circle & \Circle & \CIRCLE \\
        \CIRCLE & \Circle & \CIRCLE
    \end{pmatrix}
\end{align*}
From the left, White made a \textit{qtris} in the second row, and Black made a \textit{qtris} in the third row. Both players scored $1$ point: the game ends in a draw.
\end{enumerate}
\section{The structural analogy between QTris and QM}
In this section we describe in detail the structural analogy between QTris' rules and the mathematical framework of QM, when this latter is applied to two-state systems. 

The first aspect in which QTris' design makes contact with QM is in the general structure of the game. The three phases of the game mirror the three phases of any quantum process, experiment or computation according to the Ludwig-Kraus scheme we described above \cite{Kraus}: \emph{preparation of the state}, \emph{(unitary) transformation of the state} and \emph{measurement of an observable}. Thus, by design, every game sequence in QTris corresponds to a sequence of actual operations one can perform on quantum systems. 

The basic idea of QTris is that each cell of the grid represents a single qubit, so that the whole board represents a system of nine qubits. Thus, to fix the ideas, one can imagine that each cell represents, symbolically, a spin 1/2 systems in a Stern-Gerlach experiment, or a polarized photon in an optical circuit, or a qubit in a quantum computer. The essential feature of any two-state system, whatever may be its physical realization, is that it is described by \emph{binary}, \emph{stable} and  \emph{mutually incompatible} properties:
\begin{itemize}
    \item A property is \emph{binary} if, whenever one performs a measurement of it - i.e., one "asks a question" to the system about such property using a suitable apparatus - there are always \emph{two} possible mutually exclusive results or "answers" (\emph{et tertium non datur}). In QM binary properties are described by hermitian operators acting on a complex two-dimensional vector space. Following Susskind \cite{Susskind}, we regard hermitian operators simply as a compact way to encode the relationship between a set of real numbers - the eigenvalues of the observable - and a set of corresponding vectors - the eigenstates of the observable. As well known, from the spectral theorem it follows that the eigenstates of any observable form an orthonormal basis for the space of states. Therefore, given any observable, every state is represented in a unique way as a linear combination (or "superposition") of its eigenstates;
    \item A physical property is \emph{stable} if, for every state, whenever the \emph{same} measurement is repeated twice, the \emph{same} result is obtained \emph{with certainty}, provided nothing happens between the two measurements. In QM, stability of physical properties translates into the circumstance that every (pure) state must be the eigenstate of some observable, thus giving always the same result, with certainty, if that observable is repeatedly measured;
    \item We already described under what conditions two properties are said to be \emph{incompatibile}. We only recall that, in QM, such properties are represented by non-commuting hermitian operators acting on the state space.
    \end{itemize}
To inject these features into QTris, it is sufficient to  postulate that every cell represents a two-state system described by only \emph{two}, maximally incompatible, physical properties: \emph{Color} ($C$) and \emph{Orientation} ($O$). Color and Orientation play, in QTris, the role that any two maximally incompatible observables $A$ and $B$ play in the description of two-state systems. Thus, the state space of a QTris cell in the basic version may be conceived as the "minimal" state-space structure arising from the existence of only two (maximally) incompatible observables. If one thinks, for example, to the Stern-Gerlach setting, $A$ and $B$ may be any two orthogonal components of the spin, say $\sigma_z$ and $\sigma_x$. More abstractly, following standard use, in the following we will identify Color and Orientation with the Pauli operators $Z$ and $X$, which are both unitary and hermitian.

As quantum systems can be in different states, which encode the measurement probabilities associated to all possible observables, so in QTris a cell can be in different states represented by different tiles. As we said, in the basic version there are four possible states, and the final measurement is always a color measurement.\footnote{One may introduce also \emph{orientation measurements}, with the additional rule that all measurement probabilities must be symmetrical under exchange of preparation and outcome.} The four basic tiles in QTris are designed to suggest which property has a definite value when the cell is in that state: \emph{color-tiles}, $\Circle$ and $\CIRCLE$, represent states of a cell having definite color and maximally undetermined orientation, while \emph{orientation-tiles}, $\LEFTcircle$ and $\RIGHTcircle$, represent states of a cell having definite orientation and maximally undetermined color. In short, color-tiles and orientation-tiles are, in QTris, the "eigenstates" of Color and Orientation. Thus, in QTris, color-tiles and orientation-tiles play the same role that the basis eigenstates of any two maximally incompatible observables $A$ and $B$ play in the state space of any two-state system. If we identify Color and Orientation with the Pauli observables $Z$ and $X$ and adopt the standard notations $\{\ket0, \ket 1\}$ and $\{\ket+, \ket-\}$ for their eigenstates, in Dirac notation we may write symbolically the following formal analogy:
\begin{equation*}
    \{\ket\Circle, \ket\CIRCLE\} : \{\ket\LEFTcircle, \ket\RIGHTcircle\} = \{\ket0, \ket1\} : \{\ket+, \ket-\},
\end{equation*}
meaning that all the mutual relations between the states $\{\ket0, \ket1\}$ and $\{\ket+, \ket-\}$ are mirrored by those between the states $\{\ket\Circle, \ket\CIRCLE\}$ and $\{\ket\LEFTcircle, \ket\RIGHTcircle\}$. As well known, since states are represented by \emph{directions}, these relationship are simply \emph{angular} relationships encoded by the inner product between states.

As quantum states can be transformed into each other through reversible operations described by \emph{unitary operators} acting on the state space, in QTris tiles can be transformed using move cards. As their names suggest, the cards available in the basic version correspond to the set of Pauli operators $\{I, X, Y, Z\}$ with the addition of the \emph{Hadamard} operator  $H = (X + Z)/\sqrt{2}$, which plays a fundamental role in quantum computation. Figure \ref{fig:algebra_1} shows how in QTris the cards act on the four basic states, and it is easy to check that these rules are the same as those describing the action of Pauli operators on the corresponding $Z$ and $X$ eigenstates of a qubit. The role of $H$ is, in essence, to connect the color basis $\{\ket\Circle, \ket\CIRCLE\}$ to the orientation basis $\{\ket\LEFTcircle, \ket\RIGHTcircle\}$, or, in jargon, to "to generate a superposition of states".

Thus, in light of the above analogies, in the color basis QTris' fundamental states can be written in the usual form:
\begin{align*}
    \ket{\Circle}  &= \begin{pmatrix}
        1 \\ 0 \end{pmatrix}, \\
    \ket{\CIRCLE}  &=  \begin{pmatrix}
        0 \\ 1 \end{pmatrix}, \\
        \ket{\LEFTcircle} &= \frac{1}{\sqrt{2}}\ket{\Circle} + \frac{1}{\sqrt{2}}\ket{\CIRCLE} = \begin{pmatrix}
        \frac{1}{\sqrt{2}} \\ \frac{1}{\sqrt{2}}  \end{pmatrix}, \\
    \ket{\RIGHTcircle} &= \frac{1}{\sqrt{2}}\ket{\Circle} - \frac{1}{\sqrt{2}}\ket{\CIRCLE} = \begin{pmatrix}
        \frac{1}{\sqrt{2}}  \\ -\frac{1}{\sqrt{2}}  \end{pmatrix},
        \end{align*} 
and QTris' cards are represented by the usual matrices:
\begin{align*}
    X &= \ketbra{\Circle}{\CIRCLE} + \ketbra{\CIRCLE}{\Circle} = \begin{pmatrix} 0 & 1 \\ 1 & 0 \end{pmatrix}; \\
    Y &= -i\ketbra{\Circle}{\CIRCLE} + i\ketbra{\CIRCLE}{\Circle} =\begin{pmatrix} 0 & -i \\ i & 0 \end{pmatrix}; \\
    Z &= \ketbra{\Circle}{\Circle} - \ketbra{\CIRCLE}{\CIRCLE} =\begin{pmatrix} 1 & 0 \\ 0 & -1 \end{pmatrix}; \\ 
    H &= \frac{X+Z}{\sqrt{2}} = \frac{1}{\sqrt{2}}\begin{pmatrix} 1 & 1 \\ 1 & -1 \end{pmatrix}; \\
    I &=  \ketbra{\Circle}{\Circle} + \ketbra{\CIRCLE}{\CIRCLE} = \begin{pmatrix} 1 & 0 \\ 0 & 1 \end{pmatrix}. \\
\end{align*}
As we already remarked, the final measurement is a color-measurement, so it is natural to express all the states of QTris in the color basis, which thus plays the role of \emph{computational basis}. Using the Born Rule it is easy to recover all the probabilities given in Fig.~\ref{fig:algebra_1}.

\begin{table}[ht]
\centering
\begin{tabular}{ccc}
\toprule
Level & QTris (basic) & QM (qubits) \\
\midrule
\multirow{2}{*}{Observables}
& Color & $Z$ \\
& Orientation & $X$  \\
\multirow{2}{*}{States}
& $\{\Circle, \CIRCLE\}$ & $\{\ket0, \ket1\}$ \\
& $\{\LEFTcircle, \RIGHTcircle\}$ & $\{\ket+, \ket-\}$ \\
Unitary Operations & Move Cards & $\{I, X, Y, Z, H\}$\\
Measurement & Die Roll & Born Rule \\
\bottomrule
\end{tabular}
\caption{Structural Analogy between QTris (basic version) and QM for systems of qubits.}
\label{tab:QTris-QM}
\end{table}
So, to summarize, the structural analogy between QTris and QM for a system of qubits works at four levels (Table \ref{tab:QTris-QM}):
\begin{enumerate}
    \item \emph{Observables}. Color and Orientation are analogous to a couple of binary and maximally incompatible observables, such as $Z$ and $X$ for a qubit;
    \item \emph{States}. Color-tiles and orientation-tiles are analogous to the orthonormal eigenstates of two such observables. All the measurement probabilities shown in Fig.~\ref{fig:algebra_1} derive from this analogy. 
    \item \emph{Unitary Operations.} The transformation properties of QTris' tiles under the action of Move Cards are analogous to the transformation properties of $Z$ and $X$ eigenstates under the action of Pauli operators, with the addition of the Hadamard operator.
    \item \emph{Measurement}. The probabilities associated to the final color measurement in QTris are analogous to the probabilities associated via  \emph{Born Rule} to a $Z$ measurement on a qubit
\end{enumerate}
 Given these connections, it is clear in which sense every game sequence QTris' represents, by design, a possible quantum experiment performed on a system of qubits. Let us consider, for example, the following game sequence:
\begin{enumerate}
    \item To prepare a single cell is in the state $\Circle$;
    \item To act on this cell with cards $X$ and $H$, in this order. We get     $\Circle \xrightarrow{X} \CIRCLE\xrightarrow{H}\RIGHTcircle$;
    \item To measure the Color of the cell. Since the state is $\RIGHTcircle$, we get
  \begin{align*}
        p(\Circle|\LEFTcircle) &= \frac{1}{2}; \\
        p(\CIRCLE|\LEFTcircle) &= \frac{1}{2}; 
    \end{align*}
\end{enumerate}
This game sequence is equivalent to the following quantum process:
\begin{enumerate}
    \item To prepare a single qubit in the state $\ket{0}$;
    \item To act on the qubit with unitary operators $X$ and $H$, in this order. We get $HX\ket{0} = H\ket{1} = \ket{-}$;
    \item To measure $Z$ on the qubit. The possible results are $\ket{0}$ and $\ket{1}$, with probabilities given by Born Rule:
    \begin{align*}
        p(0|-) &= \abs{\braket{0|-}}^2 = \frac{1}{2}; \\
         p(1|-) &= \abs{\braket{1|-}}^2 = \frac{1}{2}.
    \end{align*}
\end{enumerate}
In this way, it is easy to frame QTris' game problems which are actually typical problems in QM in disguise. Here are a few examples:
\begin{itemize}
    \item A single cell is prepared in the state $x$ and acted upon by the operations $A, B, C...$. What is the final state of the cell? If a Color (Orientation) measurement is performed, what are the results? 
    \item Given a configuration of the board, compute the probabilities of all possible results of the game ("White wins", "Black wins", "draw");
    \item Given a configuration of the board, to compute the probability that player Black (White) scores (at least) $n$ points;
    \item Given a configuration of the board and a set of cards, to compute the optimal sequence of moves which maximize the winning probabilities of both players.
\end{itemize}
Examples can be multiplied at pleasure. By the fundamental analogy, every sequence of preparation, unitary transformations and measurements can be translated into Dirac notation, thus making QTris a playground to familiarize with linear algebra, probability theory and the fundamental postulates of QM.

At a higher level, from the structural analogy it follows that QTris can also be studied within the framework of 
Quantum Game Theory (QGT). QGT is an extension of Classical Game Theory (CGT) in which classical information carriers are replaced by quantum systems, and players are allowed to manipulate quantum states through a set of admissible quantum operations. Since its introduction, the field has developed into an active research area at the intersection of game theory, quantum information science and quantum computation, providing a natural framework for analyzing strategic interactions involving quantum resources such as superposition and entanglement \cite{Meyer1999,Eisert1999,Marinatto2000,BostanciWatrous2022}.

In CGT a game is described by the tuple
\begin{equation}
G=\left(N,\{S_i\}_{i\in N},\{u_i\}_{i\in N}\right),
\end{equation}
where \(N\) denotes the set of players, \(S_i\) the strategy space of player \(i\), and
\begin{equation}
u_i:S_1\times\cdots\times S_n\rightarrow\mathbb{R}
\end{equation}
the corresponding payoff function. QGT extends this framework by introducing the physical substrate on which the strategic interaction takes place. Besides the set of players and the payoff functions, a quantum game requires:
\begin{itemize}
\item a Hilbert space \(\mathcal{H}\),
\item an initial quantum state, be it represented as a density operator \(\rho\) or, for a pure state, as a vector \(\ket{\psi}\),
\item a set of admissible quantum operations, defining the players' strategies;
\item a measurement procedure determining outcome probabilities and payoffs.
\end{itemize}
The modern development of quantum game theory originated from the pioneering work of Meyer \cite{Meyer1999}, who first demonstrated that quantum strategies may outperform classical ones in suitable strategic settings. Shortly afterwards, Eisert, Wilkens and Lewenstein introduced one of the most influential protocol-based formulations of quantum games \cite{Eisert1999}, followed by the alternative approach proposed by Marinatto and Weber \cite{Marinatto2000}. Although these formulations differ in the choice of admissible strategy spaces, the role assigned to entanglement and the implementation of the game protocol, they share the same conceptual structure: strategic interactions are realized through quantum operations acting on quantum states, while payoffs emerge from measurements performed on the final state.

The general formulation of QGT is sufficiently broad to encompass strategic interactions that were not originally conceived within any of these standard protocol-based formulations. In fact, despite most of the cited works focused on simultaneous games, many realistic interactions are intrinsically sequential. In these settings, players alternately manipulate the quantum system and the state evolves through an ordered sequence of quantum operations, so that each operation depends on the previous state of the game. Sequential quantum games therefore introduce an explicit temporal structure into the strategic interaction, naturally connecting quantum game theory with quantum information processing, quantum control and quantum decision processes.

This perspective is particularly relevant for QTris, since its game structure naturally defines a strategic interaction satisfying the above criteria identifying a quantum game. In fact: 
\begin{itemize}
\item The game board represents a quantum register and it defines the Hilbert space;
\item the preparation phase corresponds to the initial quantum state;
\item the cards define the admissible quantum operations; as we will see shortly, entanglement emerges from multi-qubit operations;
\item the final measurement determines the players' payoffs. 
\end{itemize}
Therefore, QTris is a formally well defined sequential quantum game whose dynamics can be analyzed within the general framework of QGT.

To conclude this Section, let us emphasize how QTris' game design and the analogies given above put in clear light some of the key concepts described in Section 3. 

The first point, that the results of measurements in QM are in general \emph{probabilistic}, is made evident by the fact that in QTris the measurement is made by rolling a die, the random process \emph{par excellence}. The important fact that probabilities still leave room for certainty in some particular cases is exemplified by color tiles, for which the results of (color) measurements are known in advance and that, in practice, can be skipped altogether during the last phase of the game. Notice that in all the cases considered until now, the results of measurement performed on different cells are always \emph{independent}, as made evident by the fact that players roll the die once for each cell. Therefore, by definition of independent events, the probability of \emph{two} results is the \emph{product} of the probabilities of the single events. For example, if two cells are in the states $\LEFTcircle\CIRCLE$, the probabilities of obtaining under measurement the four possibilities $\Circle \Circle, \Circle \CIRCLE, \CIRCLE \Circle, \CIRCLE \CIRCLE$ are
\begin{align*}
      prob(\Circle \Circle|\LEFTcircle\CIRCLE) &= prob(\Circle|\LEFTcircle)\cdot prob(\Circle|\CIRCLE) = \frac{1}{2}\cdot 0 = 0; \\
        prob(\Circle \CIRCLE|\LEFTcircle\CIRCLE) &= prob(\Circle|\LEFTcircle)\cdot prob(\CIRCLE|\CIRCLE) = \frac{1}{2}\cdot 1 = \frac{1}{2}; \\
        prob(\CIRCLE\Circle |\LEFTcircle\CIRCLE) &= prob(\CIRCLE|\LEFTcircle)\cdot prob(\Circle|\CIRCLE) = \frac{1}{2}\cdot 0= 0; \\
         prob(\CIRCLE\CIRCLE |\LEFTcircle\CIRCLE) &= prob(\CIRCLE|\LEFTcircle)\cdot prob(\CIRCLE|\CIRCLE) = \frac{1}{2}\cdot 1= \frac{1}{2}.
\end{align*} 
As well known, this \emph{factorization} of probabilities, which naturally emerges in analyzing the possible scores in the measurement phase, is essential to distinguish the properties of \emph{product states} from those of \emph{entangled states}. This aspect will emerge clearly when we will introduce entangled states in QTris.

The second point, according to which states and the probabilities they encode can be modified in a deterministic way, is the heart of the \emph{operation phase} and, as we already emphasized, the whole strategical element in QTris depends on it. In particular, it is evident that the effects of unitary operations described in Fig.~\ref{fig:algebra_1} are perfectly known and predictable. In our experience this diagram is an extremely effective tool to understand the inner logic of QTris' operations and measurements and, therefore, the basic structure of the state space of a single qubit. Thus, putting together these two points - \emph{probabilistic} measurements and \emph{deterministic} operations - QTris' game design highlights the essential fact that quantum states differ from each other in \emph{two} essential aspects: first, they give rise in general to different probabilities under measurement; and, second, that they transform in a different way when a given unitary operator acts on them - or, what is the same, in how they transform under some specified physical conditions. 

The third point, the existence of incompatible properties, is reproduced in QTris by Color and Orientation, whose behavior under measurement replicates that of $Z$ and $X$ for systems of qubits. As we said above, this is the minimum structure that is necessary and sufficient to exhibit quantum properties, which are thus strongly connected, from the theoretical point of view, to the existence of incompatible properties.

As for the fourth point, the existence of entanglement, it will be discussed in the following Section, where we describe some extensions of QTris' basic rules.

\section{Extensions and Variants}
Given the fundamental rules of the game, QTris can be easily expanded by adding new tiles and cards corresponding to new  states and operations. In principle, there is no limit to the number of states and operations one can introduce, provided all their properties are defined without ambiguities or contradictions. Of course, adding new rules to the game also extends the class of problems that can be proposed within QTris framework. In this Section we describe some extensions of the game that add important theoretical elements of QM, such as \emph{entanglement}, \emph{partial incompatibility} and \emph{mixed states}. A further extension, the introduction of a \emph{third basis} formed by the eigenstates of $Y$, thus completing the algebra of maximally-incompatible observables of a single qubit, will be discussed in a future work.

\subsection{Entanglement in QTris: Bell states}
Entanglement in QTris is introduced by the card \emph{Controlled-X} or $C_X$, analogous to the CNOT gate of quantum computers. This is the only card of the game acting on two cells at the same time, the first playing the role of \emph{control} and the second the role of \emph{target}. The rules for the action of $C_X$ are the following: 
\begin{itemize}
    \item If the control is $\Circle$, $C_X$ acts on the target as $I$ (it does nothing). Thus we have, for example,
    \begin{align*}
         \Circle \Circle &\xleftrightarrow{C_X}  \Circle \Circle; \\
        \Circle \LEFTcircle &\xleftrightarrow{C_X} \Circle \LEFTcircle.
    \end{align*}
    \item If the control is $\CIRCLE$, $C_X$ acts on the target as $X$. Thus we have, for example,
    \begin{align*}
         \CIRCLE \Circle &\xleftrightarrow{C_X}  \CIRCLE \CIRCLE; \\
        \CIRCLE \CIRCLE &\xleftrightarrow{C_X} \CIRCLE \Circle.
    \end{align*}
    \item If the control and target are \emph{equal Orientation states}, it acts like $Z$ on both:
    \[
    \LEFTcircle\LEFTcircle \xleftrightarrow{C_X} \RIGHTcircle\RIGHTcircle.
    \]
    \item If the control is  $\LEFTcircle$ and the target is $\Circle$ or $\CIRCLE$, $C_X$ generates some new states represented by \emph{ triangular}, \emph{unsigned} tiles:
    \begin{align*}
        \LEFTcircle \Circle &\xleftrightarrow{C_X}\Triangle\Triangle; \\
        \LEFTcircle \CIRCLE &\xleftrightarrow{C_X}\Triangle\ITriangle.
    \end{align*}
    If the target is $\RIGHTcircle$, it is left unchanged.
    \item If the control is  $\RIGHTcircle$ and the target is $\Circle$ or $\CIRCLE$, $C_X$ generates some new states represented by \emph{ triangular, signed} tiles:
    \begin{align*}
        \RIGHTcircle \Circle &\xleftrightarrow{C_X}\Triangle\MTriangle; \\
        \RIGHTcircle \CIRCLE &\xleftrightarrow{C_X}\Triangle\IMTriangle;
    \end{align*}
  If the target is $\LEFTcircle$, it is left unchanged.
    \end{itemize}
So, allowing for the card $C_X$ one obtains four new possible two-cell states, called \emph{entangled states}, which are represented by \emph{couples} of signed triangular tiles $\Triangle \Triangle, \Triangle \ITriangle,  \Triangle \IMTriangle, \Triangle \MTriangle$. The triangle notation is suggestive of two important aspects. First, these states are characterized by two \emph{global} properties: \emph{Alignment}, triangles being either \emph{similarly} or \emph{oppositely} oriented, and \emph{Sign}, the second triangle being either \emph{signed} or \emph{unsigned}; second, the appearance of the tiles says nothing about the Color of the two cells, which, as we will see, is in fact completely unpredictable. Notice, in particular, that using $C_X$ on triangular states one can \emph{disentangle} them and go back to couples of circular tiles, of which one is in a Color state and the other in an Orientation state.\footnote{When a player disentangles two cell using $C_X$ he/she must still choose which cell acts as control. This choice determines the arrangement of the resulting circular tiles on two cells, as can be seen easily using the formulae given below. So, in short, when a player disentangles two cells he/she is free to choose how to arrange the resulting circular tiles. Therefore, if used twice inverting the role of control and target, $C_X$ can be usefully employed to \emph{swap} the position of two tiles on the board.}

\begin{figure}
    \centering
    \includegraphics[width = 0.65\textwidth]{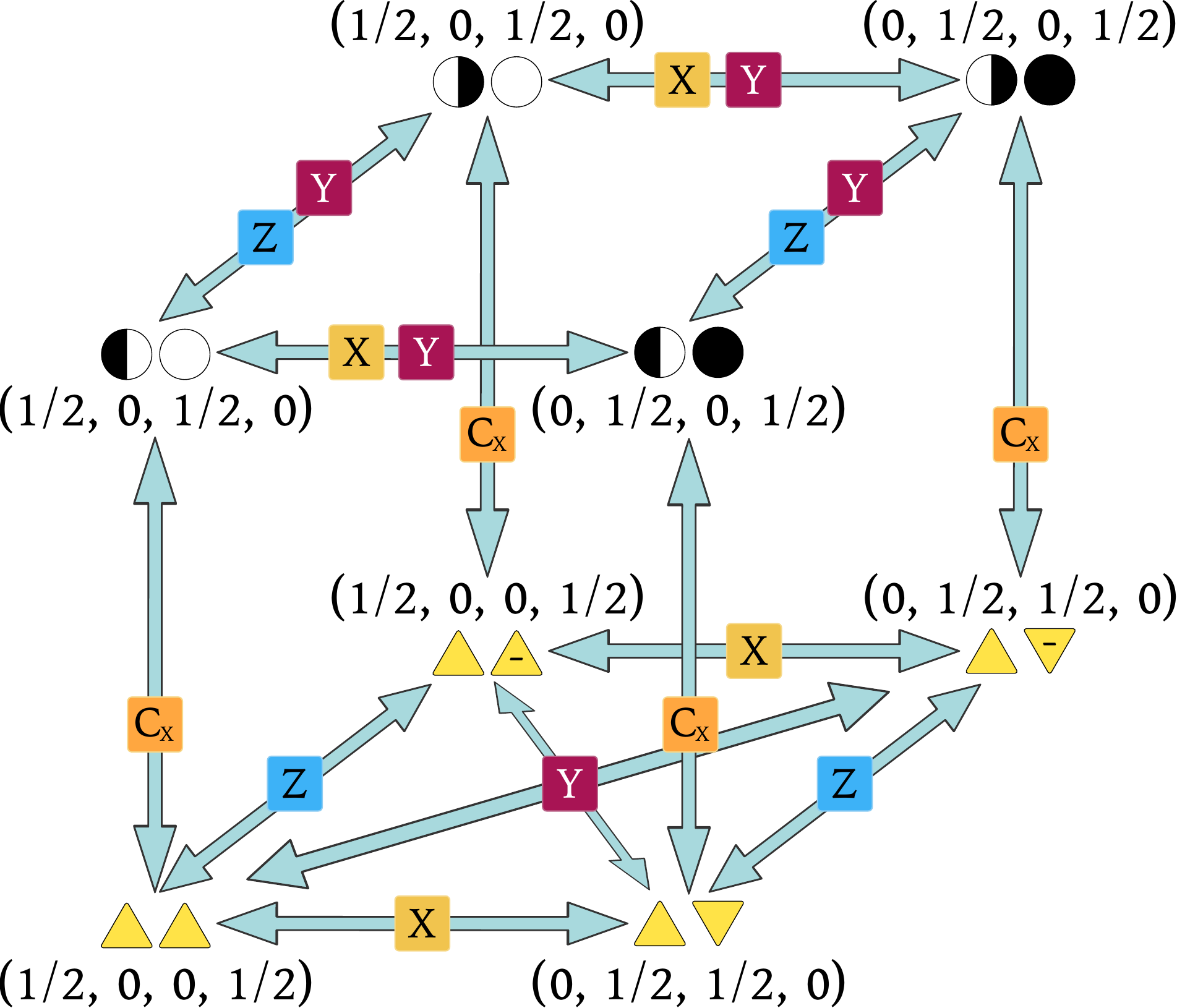}
    \caption{\textbf{Cubic map of double-cell operations $X, Y,Z, C_X$.} The numbers in the parentheses $(w, x,y,z)$, next to the symbols $ \LEFTcircle \Circle, \LEFTcircle \CIRCLE, \RIGHTcircle \CIRCLE, \RIGHTcircle \Circle$ and $\protect\Triangle \protect\Triangle, \protect\Triangle \protect\ITriangle, \protect \Triangle \protect\IMTriangle, \protect\Triangle \protect\MTriangle$, represent the probabilities of obtaining $\Circle \Circle, \Circle \CIRCLE, \CIRCLE \Circle, \CIRCLE \CIRCLE$ respectively, after performing a measurement. Two symbols on the same arrow mean that both operations achieve the same effect.}
    \label{fig:cubo_1}
\end{figure}

To fully define the meaning of these new entangled states, we must define how they behave under operations and Color measurements. The rules for measurements on triangular states are the following:
\begin{itemize}
    \item If the state is $\Triangle\Triangle$ or $\Triangle\MTriangle$, the possible results are $\Circle\Circle$ or $\CIRCLE\CIRCLE$ with probability 1/2;
    \item If the state is $\Triangle\ITriangle$ or $\Triangle\IMTriangle$, the possible results are $\Circle\CIRCLE$ or $\CIRCLE\Circle$ with probability 1/2;
\end{itemize}
As for operations on triangular states, the transformation rules for the actions of $X, Y, Z$ are the following, being indifferent on which of the two cells the card is played:
\begin{itemize}
    \item $X$ changes the \emph{Alignment} and leaves the \emph{Sign} unchanged:
    \begin{align*}
        \Triangle\Triangle \xleftrightarrow{X} \Triangle\ITriangle;  \\
          \Triangle\MTriangle \xleftrightarrow{X} \Triangle\IMTriangle.
    \end{align*}
    \item $Z$ changes the \emph{Sign} and leaves \emph{Alignment} unchanged:
    \begin{align*}
        \Triangle\Triangle \xleftrightarrow{Z} \Triangle\MTriangle;  \\
          \Triangle\ITriangle \xleftrightarrow{Z} \Triangle\IMTriangle.
    \end{align*}
    \item $Y$ changes both \emph{Alignment} and \emph{Sign}:
    \begin{align*}
        \Triangle\Triangle \xleftrightarrow{Y} \Triangle\IMTriangle;  \\
          \Triangle\ITriangle \xleftrightarrow{Y} \Triangle\MTriangle.
    \end{align*}
\end{itemize} 
The rules for generating triangular states from circular ones through $C_X$, the action of cards $X, Y, Z$ and the measurement probabilities are given in Figure \ref{fig:cubo_1}.

Before going on let us make an example of measurement on a board including entangled states. Consider the following configuration at the end of the operations phase:
\[
      \begin{pmatrix}
        \LEFTcircle & \RIGHTcircle & \Circle \\
        \CIRCLE & \CIRCLE & \Circle \\
        \Triangle & \RIGHTcircle & \ITriangle
    \end{pmatrix}.    
\]
Let us measure the tiles on the board which are not Color states using a d100:
\begin{itemize}
    \item On cell $1$, we get $54$; then $\LEFTcircle \xrightarrow{\Ameter} \CIRCLE$;
    \item On cell $2$, we get $71$; then $\RIGHTcircle \xrightarrow{\Ameter} \CIRCLE$;
    \item On cell $7$, which is entangled with $9$, we get a $99$; then for cells $7$ and $9$ we have $\Triangle \ITriangle \xrightarrow{\Ameter} \CIRCLE \Circle$;
    \item On cell $8$, we get 24; then $\LEFTcircle \xrightarrow{\Ameter} \Circle$.
    \end{itemize}
 Thus, at the end of the measurement phase, the game board is: 
\[
\begin{pmatrix}
            \CIRCLE & \CIRCLE &\Circle\\
       \CIRCLE &\CIRCLE &\Circle \\
      \CIRCLE   &\Circle &\Circle
\end{pmatrix},
\]
and both players made one \emph{qtris}: the game ends in a draw.
    
So, to summarize, whenever two cells are in a triangular state, the following conditions hold: 
\begin{enumerate}
    \item The results of a measurement on a \emph{single} cell or sub-system will be completely random. In other terms, in all cases the Color of any sub-system can be either black or white with equal probability, so that we can never predict the Color of a single cell;\footnote{In this sense, entanglement \emph{alone} does not give any real advantage in the game, as can be proven within the framework of QGT.}
    \item However, the results of the measurements on the \emph{two} cells will be \emph{not independent}. If the triangles are \emph{aligned}, the results will be \emph{equal} or (maximally) \emph{correlated} ($\Circle \Circle$ or $\CIRCLE \CIRCLE$); if the triangles are \emph{anti-aligned} the results will be \emph{opposite} or (maximally) \emph{anticorrelated} ($\Circle \CIRCLE$ or $\CIRCLE \Circle$). Notice, in particular, that if the state is entangled, probabilities for measurements performed on the sub-systems \emph{do not factorize};
    \item In the operation phase, using the old cards $\{X, Y, Z\}$ one can transform any triangular state into any other, according to the rules shown in Fig.\ref{fig:cubo_1}. In particular, disentangling a couple with $C_X$ and using the other cards ($H$ included) one can arrive at any possible couple of states.
    \end{enumerate}
The states $\Triangle \Triangle, \Triangle \ITriangle,  \Triangle \IMTriangle, \Triangle \MTriangle$ are evidently the QTris version of \emph{Bell states}. In fact, taking into account the analogies of Table \ref{tab:QTris-QM}, in the Color basis we can represent the card $C_X$ as the usual unitary operator acting on the tensor product of the state spaces of the two cells: 
\[
C_X = \ketbra{\Circle\Circle}{\Circle\Circle} + \ketbra{\Circle\CIRCLE}{\Circle\CIRCLE} + \ketbra{\CIRCLE\CIRCLE}{\CIRCLE\Circle} + \ketbra{\CIRCLE\Circle}{\CIRCLE\CIRCLE} =  \begin{pmatrix}
    1 & 0 & 0 & 0 \\
    0 & 1 & 0 & 0 \\
    0 & 0 & 0 & 1 \\
    0 & 0 & 1 & 0 \\
\end{pmatrix}.
\]
Thus, for triangular states we obtain the usual expressions of \emph{Bell states}:
\begin{align*}
   \ket{\Triangle\Triangle} &= \frac{1}{\sqrt{2}} \Bigl( \ket{\Circle\Circle} + \ket{\CIRCLE \CIRCLE}\Bigr) = \begin{pmatrix}
        \frac{1}{\sqrt{2}}  \\ 0 \\ 0 \\ \frac{1}{\sqrt{2}}  \end{pmatrix};  \\
   \ket{\Triangle\MTriangle} &= \frac{1}{\sqrt{2}} \Bigl( \ket{\Circle\Circle} - \ket{\CIRCLE \CIRCLE}\Bigr) = \begin{pmatrix}
        \frac{1}{\sqrt{2}}  \\ 0 \\ 0 \\ -\frac{1}{\sqrt{2}}  \end{pmatrix}; \\
   \ket{\Triangle\ITriangle} &= \frac{1}{\sqrt{2}} \Bigl( \ket{\Circle\CIRCLE} + \ket{\CIRCLE \Circle}\Bigr) = \begin{pmatrix}
         0 \\ \frac{1}{\sqrt{2}}  \\ \frac{1}{\sqrt{2}} \\0  \end{pmatrix}; \\
   \ket{\Triangle\IMTriangle} &= \frac{1}{\sqrt{2}} \Bigl( \ket{\Circle\CIRCLE} - \ket{\CIRCLE \Circle}\Bigr) = \begin{pmatrix}
         0 \\ \frac{1}{\sqrt{2}}  \\ -\frac{1}{\sqrt{2}} \\0  \end{pmatrix}.
\end{align*}
It is easy to verify that all the measurement probabilities given in Fig.~\ref{fig:cubo_1} are obtained from the above expressions through Born Rule.

As for the operations, we have here the first of a few game design choices made in order to improve QTris' playability. In fact, it is evident that two-cell states are acted upon by operators that, mathematically speaking, are different from those acting on single-cell states, since they act in different spaces. Nevertheless, in QTris we chose to use the same cards to represent unitary operators acting on both single and entangled systems. The simple rule behind the transformations given above is that every time a card is used on a cell in a triangular state, it is implied that the card $I$ is played on the other. In other terms, with the exception of $C_X$, all other cards in QTris represent \emph{local} operations acting on a sub-system leaving unchanged the other one. It is an easy exercise to verify that, for Bell states, if $P$ is any local Pauli operator, the effect of $I\otimes P$ is equal to the effect of $P\otimes I$. This is also why in the game it is indifferent on which of the two cells in a triangle state one plays a card.

To close this Section, let us emphasize how QTris' game mechanics for triangular states provides a simple way to bypass the need for semi-classical analogies and understand some key aspects of entanglement in a clear, non-ambiguous, operational way:
\begin{enumerate}
    \item As usual in quantum computation, entangled states are generated in QTris by the successive use of $H$ and $C_X$ on a couple of $Z$ eigenstates. The same operations in opposite order are sufficient to disentangle a given couple and obtain a couple of $Z$ eigenstates in on which $Z$ measurements give certain results. Thus, the very \emph{existence} of entangled states is immediately connected (1) to the existence of incompatible properties, which, in turn, is connected to the existence of an operation $H$ connecting the basis vectors associated to such properties; and (2) to some \emph{interaction} occurred in the past between the two systems, as the (classical) CNOT operation effected by $C_X$. Thus, in particular, incompatible properties are seen to be a \emph{necessary} (but not sufficient) condition for entanglement: in \emph{this} sense, one can say that entanglement is a "purely quantum" phenomenon. Moreover, there is no question of "spooky action at a distance": as always, the measurement probabilities depend on how the two systems have been previously prepared.
    \item When a measurement is performed on sub-systems that are part of an entangled system, the distributions of results of these measurement are not independent and show correlations following well-defined laws. In QTris' game mechanics this non-independency is highlighted very simply by the fact that, if two cells are in a triangular state, to measure the Color of \emph{both} cells one has to roll the die only \emph{once}. However, as we already emphasized, the probabilities of the two possible results are equal, so that one can never predict in advance the Color of a single sub-system.\footnote{Formally, this corresponds to the fact that the partial trace of Bell state is the completely mixed state.}  In \emph{this} sense, one can say (following Susskind \cite{Susskind}) that, for entangled states, despite we know "everything" about the \emph{global} properties of system (i.e., correlation or anticorrelation between the Colors of the cells), we know "nothing" about the \emph{local} properties of the single sub-systems (i.e., the Color of the single cells). This odd property of entangled systems is embedded in QTris' measurement rules for triangular states.
    \item Like all quantum states, also entangled states can be transformed in a deterministic way through unitary operations. Therefore, like probabilities, also correlations arising from pure entangled states can be modified at will. In particular, signed and unsigned triangular states behave identically under measurement, but differ in how they transform under unitary operations. In this regard, Figure \ref{fig:cubo_1} plays for entangled states the same role that Figure \ref{fig:algebra_1} played for single cells: it is a powerful tool to understand the mutual relationships of Bell states, their relationship with separable states, and the measurement correlations they encode. 
\end{enumerate}
\subsection{Non-flat Probabilities and Partial Incompatibility}
The probability distributions encoded in the states considered until now are such that all the probabilities different from $0$ and $1$ are equal among themselves (e.g., $(\frac{1}{2}, \frac{1}{2}), (\frac{1}{2}, 0, 0, \frac{1}{2}), (0, \frac{1}{2}, \frac{1}{2}, 0)$); in short, such probability distributions are \emph{flat}. It is easy to see that, if one considers a measurement on a single cell in such a state, players can have no advantage with respect to each other, since the result will be completely random. Thus, in these situations, the only  advantage the players can have comes from the global configuration of the board. However, to explore the consequences of non-flat probabilities in QM and, at the same time, to make QTris more interesting, we may add the card $U$.\footnote{For the moment, let us exclude the card $C_X$. We will consider $U$ and $C_X$ together in a following subsection.}

\begin{figure*}
\centering\includegraphics[width=0.5\textwidth]{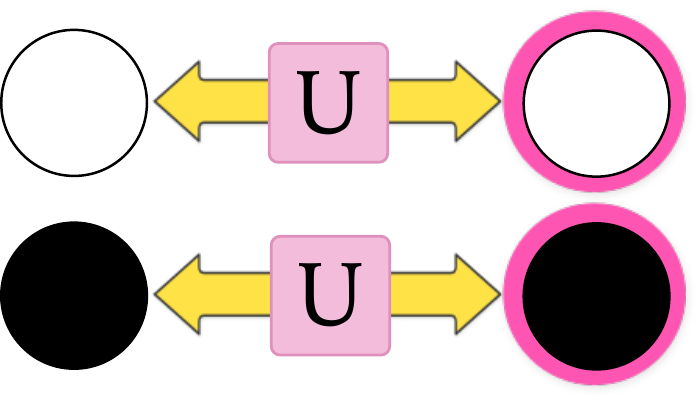}
    \caption{
    \textbf{Effect of the $U$ operation card on single-cell tiles.} After applying the $U$ card to $\Circle, \CIRCLE, \LEFTcircle, \RIGHTcircle$ you must decorate it by placing a pink tile underneath. Conversely, if you apply the $U$ card again, you must remove the pink tile. This decoration rule is also valid when using a $U$ card on any tile of an entangled state, see below.}
    \label{fig:azione_U}
\end{figure*}

The effect of the card $U$ on the four basic circular tiles is to \emph{decorate} them with a pink tile, so to obtain what we will call a \emph{$U$-decorated state}; conversely, one can remove the pink decoration by acting with $U$ on a $U$-decorated tile (Fig. \ref{fig:azione_U}):
\begin{align*}
    \Circle &\xleftrightarrow{U} \UTileCircle \\
    \CIRCLE &\xleftrightarrow{U} \UTileCIRCLE \\
    \LEFTcircle &\xleftrightarrow{U} \UTileLcircle \\
    \RIGHTcircle &\xleftrightarrow{U} \UTileRcircle 
\end{align*}
So, in short, allowing for the card $U$ (but excluding the card $C_X$) introduces four new $U$-decorated states.

\begin{figure} 
\centering\includegraphics[width=0.5\textwidth]{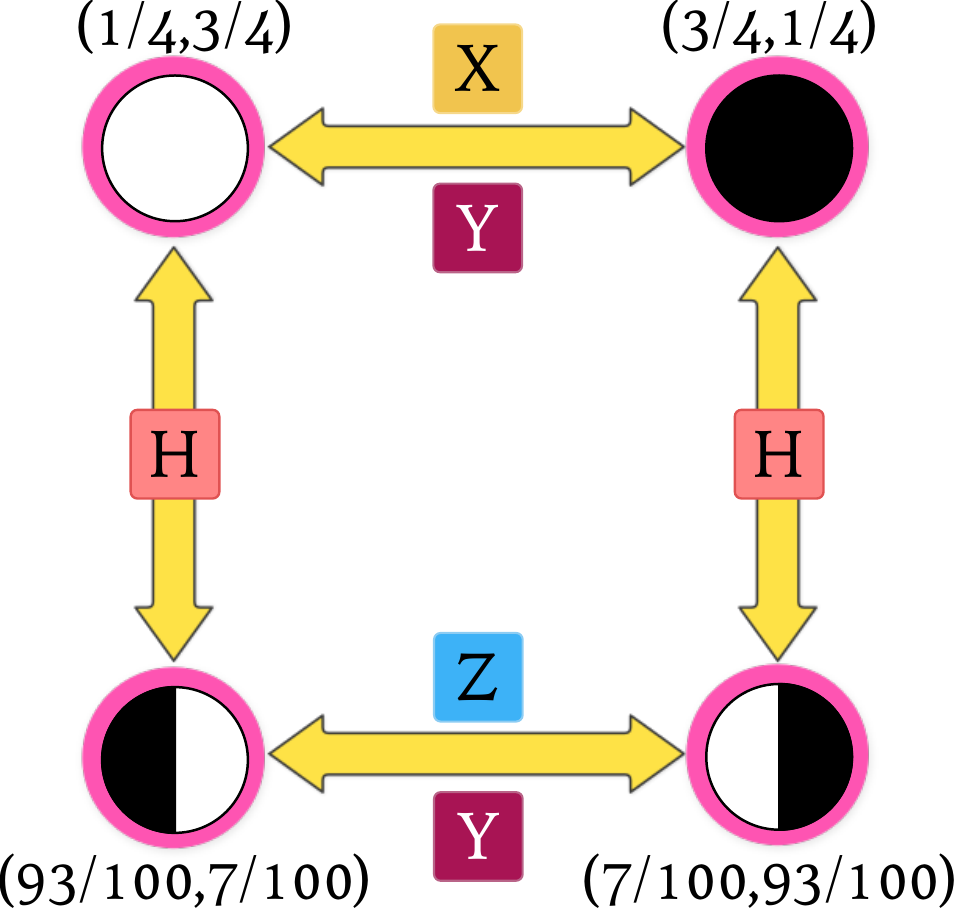}
    \caption{\textbf{Map of single-cell operations $X, Y, Z, H$ on the decorated $U$ tiles.} 
    The numbers in the parentheses $(x,y)$ next to the symbols $\protect \UTileCircle, \protect\UTileCIRCLE, \protect\UTileLcircle, \protect\UTileRcircle$ represent the probabilities of obtaining $\Circle$ and $\CIRCLE$ respectively, after performing a measurement. Two symbols on the same arrow mean that both operations achieve the same effect.}
    \label{fig:algebra_U}
\end{figure}

As usual, to define the meaning of these new states we must define how they behave under operations and measurements. Under measurements, $U$-decorated states behave as follows:
\begin{itemize}
    \item If the state is $\UTileCircle$, the probabilities for $\Circle$ and $\CIRCLE$ are $\frac{1}{4}$ and $\frac{3}{4}$;
    \item If the state is $\UTileCIRCLE$, the probabilities for $\Circle$ and $\CIRCLE$ are $\frac{3}{4}$ and $\frac{1}{4}$;
    \item If the state is $\UTileLcircle$, the probabilities for $\Circle$ and $\CIRCLE$ are $\frac{93}{100}$ and $\frac{7}{100}$;
    \item If the state is $\UTileRcircle$, the probabilities for $\Circle$ and $\CIRCLE$ are $\frac{7}{100}$ and $\frac{93}{100}$.
\end{itemize}
As for operations, cards $I, X, Y, Z, H$ act between $U$-decorated tiles as they do between non-decorated ones. Thus, for example, we have
\begin{align*}
    \UTileCircle &\xleftrightarrow{X} \UTileCIRCLE;\\
     \UTileLcircle &\xleftrightarrow{Z} \UTileRcircle;\\
    \UTileCircle &\xleftrightarrow{H} \UTileLcircle.
\end{align*}
All the measurement probabilities and the transformation rules of $U$-decorated states are given in Figure \ref{fig:algebra_U}.

Thanks to these new states QTris' game mechanics also displays the effects of \emph{partial incompatibility}. In fact, as all possible states, also $U$-decorated states must be the eigenstates of some observable. Let us call \emph{$U$-color} the observable associated to the states $\{\UTileCircle, \UTileCIRCLE\}$ and \emph{$U$-orientation} the observable associated to the states $\{\UTileLcircle, \UTileRcircle\}$. Now we ask: is the property $U$-color compatible with Color or not? Take, for example, the state $\UTileCircle$. The measurement rules give:
\begin{align*}
    prob(\Circle|\UTileCircle) &= \frac{1}{4} \\
    prob(\CIRCLE|\UTileCircle) &= \frac{3}{4} \\
\end{align*}
Thus, if the system has a definite value for $U$-color, its Color is indeterminate, but \emph{not completely}: in fact, the probability of getting $\CIRCLE$ is three times that of getting $\Circle$, and this state will clearly give a strategic advantage to the Black player. Therefore, whenever we know the $U$-color of a tile, we \emph{do} have \emph{some} information about its Color, and \emph{viceversa}: this is what \emph{partial incompatibility} is. The same happens with $U$-orientation: in the case of $\UTileLcircle$, for example, we have:
\begin{align*}
   prob(\Circle|\UTileLcircle) &= \frac{93}{100} \\
    prob(\CIRCLE|\UTileLcircle) &= \frac{7}{100};
\end{align*}
thus, if a cell is in the state $\UTileLcircle$, its Color is still indeterminate, but we are \emph{almost certain} that after measurement the result $\Circle$ will be observed - which of course will give a great advantage to the White player.

To formalize the card $U$ in the language of linear algebra it is sufficient to remember that any unitary operation on a single qubit can be written as a combination of ket-bras of the computational basis in the following way:
\[
U (\theta, \phi) = \cos\theta\ketbra{0}{0} + e^{-i \phi} \sin\theta \ketbra{0}{1} + e^{i \phi} \sin\theta \ketbra{1}{0} - \cos\theta \ketbra{1}{1}
\]
where $\theta\in[0, \pi]$ and $\phi\in[0, 2\pi]$ are real parameters. The operator corresponding to the card $U$ in QTris is obtained by choosing
$\phi = 0$ and $\theta = \pi/3$, so that in the computational basis we have
\[
U = \frac{1}{2}  \ketbra{0}{0} + \frac{\sqrt{3}}{2}  \ketbra{0}{1} + \frac{\sqrt{3}}{2}\ketbra{1}{0} - \frac{1}{2} \ketbra{1}{1} = \begin{pmatrix} \frac{1}{2} & \frac{\sqrt{3}}{2} \\ \frac{\sqrt{3}}{2} & -\frac{1}{2} 
\end{pmatrix},
\]
which, following the analogy of Table \ref{tab:QTris-QM}, in QTris-notation becomes
\[
U = \frac{1}{2}  \ketbra{\Circle}{\Circle} + \frac{\sqrt{3}}{2}  \ketbra{\Circle}{\CIRCLE} + \frac{\sqrt{3}}{2}\ketbra{\CIRCLE}{\Circle} - \frac{1}{2} \ketbra{\CIRCLE}{\CIRCLE} = \begin{pmatrix} \frac{1}{2} & \frac{\sqrt{3}}{2} \\ \frac{\sqrt{3}}{2} & -\frac{1}{2} 
\end{pmatrix}.
\]
Using this expression for $U$ we obtain the following expression for $U$-decorated states,
\begin{align*}
    \ket{\UTileCircle} &= \frac{1}{2} \ket{\Circle} +   \frac{\sqrt{3}}{2} \ket{\CIRCLE},\\
    \ket{\UTileCIRCLE} &= \frac{\sqrt{3}}{2}  \ket{\Circle} -  \frac{1}{2} \ket{\CIRCLE}, \\
    \ket{\UTileLcircle} &= \frac{\sqrt{3} + 1}{2 \sqrt{2}} \ket{\Circle} + \frac{\sqrt{3}-1}{2 \sqrt{2}} \ket{\CIRCLE}, \\
    \ket{\UTileRcircle} &= \frac{1- \sqrt{3}}{2 \sqrt{2}}  \ket{\Circle} -\frac{1+ \sqrt{3}}{2 \sqrt{2}} \ket{\CIRCLE},
\end{align*}
and using Born Rule it is easy to recover all the measurement probabilities given in Fig.~\ref{fig:algebra_U}. For example, if a cell is in the state $\UTileLcircle$, we have the Color probabilities:
\begin{align*}
    prob(\Circle|\UTileLcircle) &= \abs{\braket{\Circle|\UTileLcircle}}^2 = \abs{\frac{\sqrt{3} + 1}{2 \sqrt{2}}}^2 \simeq 0.93 \\
    prob(\CIRCLE|\UTileLcircle) &= \abs{\braket{\CIRCLE|\UTileLcircle}}^2 = \abs{\frac{\sqrt{3}-1}{2 \sqrt{2}}}^2 \simeq 0.07.
\end{align*}

For the action of the cards $X, Y, Z, H$ on $U$-decorated tiles a few comments are in order. In fact, using the expressions given above, it is easy to see that the rules given in Fig.~\ref{fig:algebra_U} are not satisfied. We have, for example,
\[
X\ket{\UTileCircle} = \frac{1}{2} X\ket{\Circle} +   \frac{\sqrt{3}}{2} X\ket{\CIRCLE} = \frac{1}{2} \ket{\CIRCLE} +   \frac{\sqrt{3}}{2} \ket{\Circle} \neq \ket{\UTileCIRCLE}.
\]
In other terms, the operator which transforms $\UTileCircle$ into $\UTileCIRCLE$ is \emph{not} the same operator which transforms $\Circle$ into $\CIRCLE$. It is easy to verify that the transformation between $\UTileCircle$ and $\UTileCIRCLE$ is effected by the operator
\[
X_U = UXU^\dagger,
\]
since we have
\[
X_U\ket{\UTileCircle} = UXU^\dagger U\ket{\Circle} = UX\ket{\Circle} = U\ket\CIRCLE = \ket{\UTileCIRCLE}.
\]
Along the same lines it is easy to prove that, given any linear operator $P$ and any unitary $U$, if $\ket{b} = P\ket{a}$, $U\ket{a} = \ket{a'}$, and $U\ket{b} = \ket{b'}$, then $\ket{b'} = UPU^\dagger \ket{a'}$. In other terms: the operator $U$ which transforms the states also transforms the operator connecting them. This is why, in QTris, we chose to use the same cards $X, Y, Z, H$ connecting the non-decorated states to represent \emph{also} the U-transformed operators $UXU^\dagger, UYU^\dagger, UZU^\dagger, UHU^\dagger$, connecting the $U$-decorated states. In this way, a twofold aim is achieved: first, without introducing new cards we avoid over-complexity and preserve QTris' playability also for this extended scenario; second, in this way we have the opportunity to emphasize the essential property of unitary operations as those preserving the inner product between any two states, and, therefore, that \emph{do not alter their mutual relationship}. In fact, using the expressions given above, it is easy to verify that $U$-decorated states are connected to each other by equations that are identical to those relating non-decorated Color and Orientation states:
\begin{align*}
    \ket{\UTileLcircle} &= \frac{1}{\sqrt{2}}\ket{\UTileCircle} + \frac{1}{\sqrt{2}}\ket{\UTileCIRCLE} \\
    \ket{\UTileRcircle} &= \frac{1}{\sqrt{2}}\ket{\UTileCircle} - \frac{1}{\sqrt{2}}\ket{\UTileCIRCLE}.
\end{align*} 
In particular, it follows that $U$-color and $U$-orientation are \emph{maximally incompatible}, just like Color and Orientation.

\begin{figure} 
    \centering
    \includegraphics[width=\textwidth]{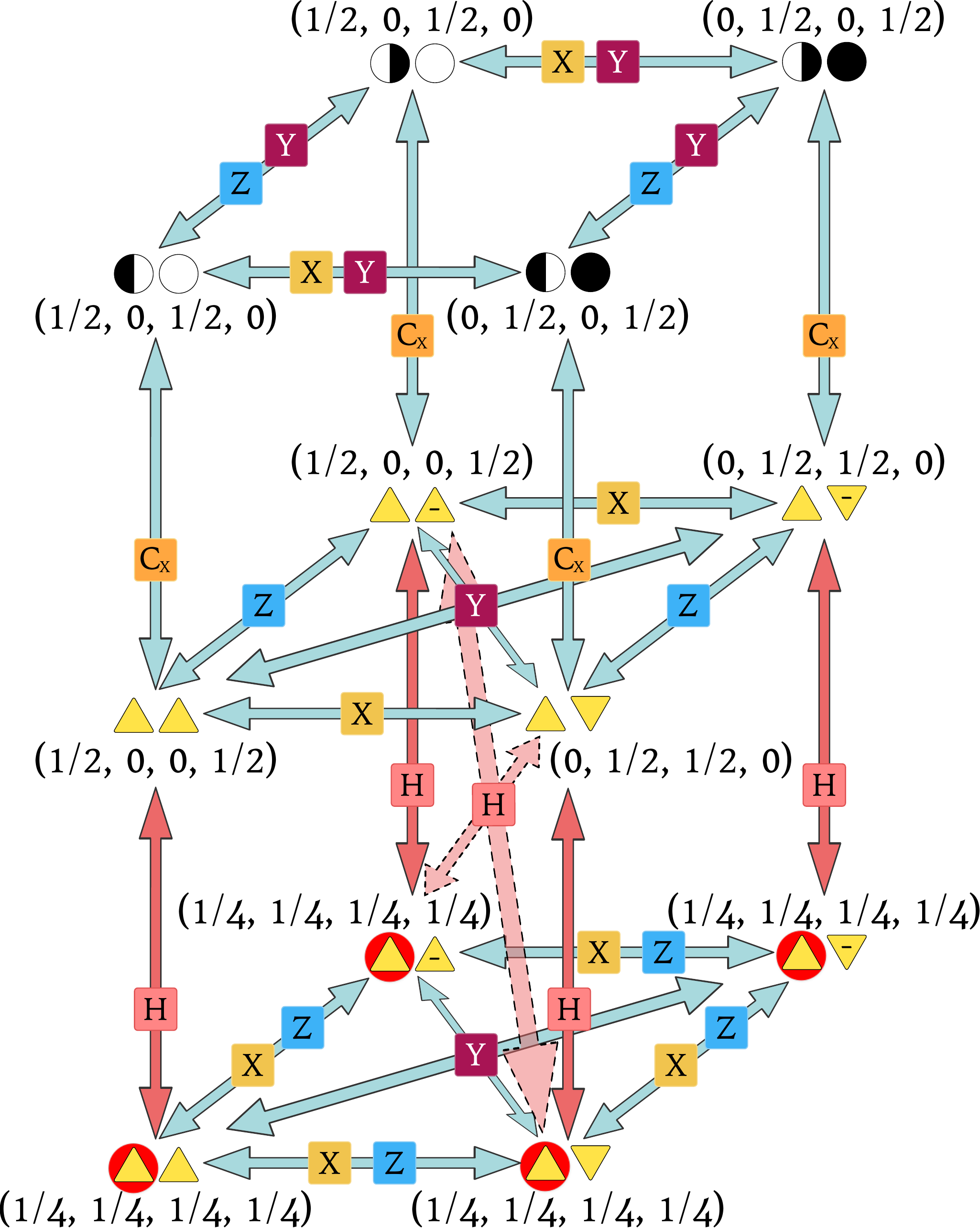}
    \caption{\textbf{Map of all the operations allowed by using entangled states together with $H$ and the corresponding probabilities for the outcomes $\{ \Circle \Circle, \Circle \CIRCLE, \CIRCLE \Circle, \CIRCLE \CIRCLE \}$.} Two symbols on the same arrow mean that both operations achieve the same effect.}
    \label{fig:ParallelepipedoH}
\end{figure}
\subsection{More Entangled States}
What happens when we use $H$ and $U$ on entangled states? To answer this question, we only have to use the above expressions for the operators and keep in mind our previous remarks about \emph{local} operations: every time we use a card on an entangled couple we must choose on which sub-system it acts. For Bell states and Pauli operators the effect was independent from this choice, but, as we will see shortly, in general this is not the case.

Let us begin with $H$ (see Fig.~\ref{fig:ParallelepipedoH}). When we use $H$ on a Bell couple we must add a red tile underneath the upper unsigned triangle, thus obtaining one of four possible \emph{$H$-decorated} entangled states:
\[
    \{ \Triangle \Triangle, \Triangle \ITriangle, \Triangle \MTriangle, \Triangle \IMTriangle \} \xrightarrow{H} \{ \HTileTriangle \Triangle, \HTileTriangle \ITriangle, \HTileTriangle \MTriangle, \HTileTriangle \IMTriangle \}.
\]
Using the above expressions for $H$ and for Bell states, in the color basis the four new $H$-decorated entangled states are written as follows:
\begin{equation}
\begin{aligned}
    \ket{\HTileTriangle \Triangle} &= \frac{1}{2}\Bigl(\ket{\Circle \Circle} + \ket{\Circle\CIRCLE} + \ket{\CIRCLE\Circle} - \ket{\CIRCLE\CIRCLE}\Bigr) \\
    \ket{\HTileTriangle \ITriangle} &= \frac{1}{2}\Bigl(\ket{\Circle \Circle} + \ket{\Circle\CIRCLE} - \ket{\CIRCLE\Circle} + \ket{\CIRCLE\CIRCLE}\Bigr) \\
    \ket{\HTileTriangle \MTriangle} &= \frac{1}{2}\Bigl(\ket{\Circle \Circle} - \ket{\Circle\CIRCLE} + \ket{\CIRCLE\Circle} + \ket{\CIRCLE\CIRCLE}\Bigr) \\
    \ket{\HTileTriangle \IMTriangle} &= \frac{1}{2}\Bigl(-\ket{\Circle \Circle} + \ket{\Circle\CIRCLE} + \ket{\CIRCLE\Circle} + \ket{\CIRCLE\CIRCLE}\Bigr).
\end{aligned}
\end{equation}
Using the usual Born Rule it is easy to verify that the measurement probabilities of these new states are flat and all equal to $\frac{1}{4}$, as shown in Fig.~\ref{fig:ParallelepipedoH}. Notice, however, that if the initial state is either $\Triangle \ITriangle$ or $\Triangle \MTriangle$, using $H$ we can also choose to move along diagonal directions, obtaining the alternative transformations
\[
\{ \Triangle \ITriangle, \Triangle \MTriangle \} \xrightarrow{H} \{ \HTileTriangle\MTriangle, \HTileTriangle\ITriangle  \};
\]
in other terms, for the first time using a card on a state can produce two distinct effects corresponding to the choice of the cell on which the $H$ card operates.  In fact, it is easy to verify that the vertical arrows represent the effect of $H\otimes I$, while the diagonal arrows represent the effect of $I\otimes H$.\footnote{In practice, during the game it is not necessary to specify on which qubit one applies the $H$ card, but just to specify which transformation one wants to perform.} For example, we have
\begin{align*}
    H\otimes I \ket{\Triangle \ITriangle} &= \frac{1}{\sqrt{2}} \Bigl( \ket{\LEFTcircle\CIRCLE} + \ket{\RIGHTcircle \Circle}\Bigr) = \frac{1}{2}\Bigl(\ket{\Circle\CIRCLE} + \ket{\CIRCLE\CIRCLE} + \ket{\Circle \Circle} - \ket{\CIRCLE\Circle}\Bigr) = \ket{\HTileTriangle \ITriangle}\\
    I\otimes H \ket{\Triangle \ITriangle} &= \frac{1}{\sqrt{2}} \Bigl( \ket{\Circle\RIGHTcircle} + \ket{\CIRCLE \LEFTcircle}\Bigr) =  \frac{1}{2}\Bigl(\ket{\Circle \Circle} - \ket{\Circle\CIRCLE} + \ket{\CIRCLE\Circle} + \ket{\CIRCLE\CIRCLE}\Bigr) =  \ket{\HTileTriangle \MTriangle}. 
\end{align*}
As indicated in the bottom face of Fig.~\ref{fig:ParallelepipedoH}, the same holds for $X$ and $Z$, which can change either Alignment or Sign of the couple according to the sub-system on which they act. For example, we have
\begin{align*}
X\otimes I \ket{\HTileTriangle \Triangle} &= \frac{1}{2}\Bigl(\ket{\CIRCLE \Circle} + \ket{\CIRCLE\CIRCLE} + \ket{\Circle\Circle} - \ket{\Circle\CIRCLE}\Bigr) =     \ket{\HTileTriangle \MTriangle}; \\
I\otimes X \ket{\HTileTriangle \Triangle} &= \frac{1}{2}\Bigl(\ket{\Circle \CIRCLE} + \ket{\Circle\Circle} + \ket{\CIRCLE\CIRCLE} - \ket{\CIRCLE\Circle}\Bigr) = \ket{\HTileTriangle \ITriangle}. \\
\end{align*}

\begin{figure*}
    \centering
\includegraphics[width=0.95\textwidth]{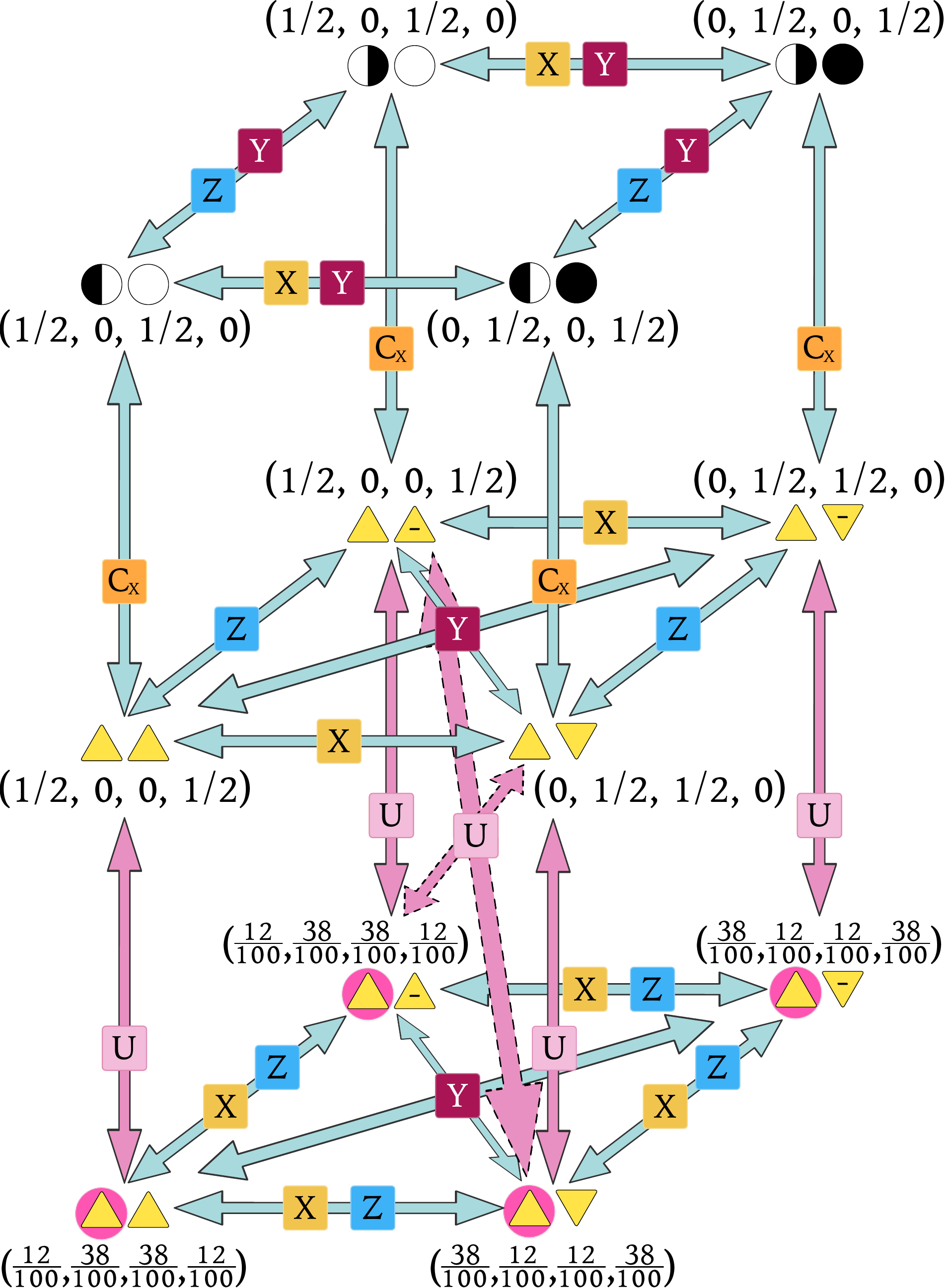} 
    \caption{\textbf{Map of all the operations allowed by using entangled states together with $U$ and the corresponding probabilities for the outcomes $\{ \Circle \Circle, \Circle \CIRCLE, \CIRCLE \Circle, \CIRCLE \CIRCLE \}$.} Two symbols on the same arrow mean that both operations achieve the same effect. } \label{fig:ParallelepipedoU}
\end{figure*} 

Exactly in the same way we can define the action of the card $U$ on entangled states (see Fig.~\ref{fig:ParallelepipedoU}). When we use $U$ on a triangle couple we must add a pink tile underneath the upper unsigned triangle, thus obtaining one of the four \emph{$U$-decorated} entangled states: 
\[
    \{ \Triangle \Triangle, \Triangle \ITriangle, \Triangle \MTriangle, \Triangle \IMTriangle \} \xrightarrow{U} \{ \UTileTriangle \Triangle, \UTileTriangle \ITriangle, \UTileTriangle \MTriangle, \UTileTriangle \IMTriangle \}.
\]
These transformations correspond to the vertical lines in Fig.~\ref{fig:ParallelepipedoU}. Again, and for the same reasons as before, if the initial state is either $\Triangle \ITriangle$ or $\Triangle \MTriangle$, using $U$ we can also move along diagonal directions, obtaining the alternative transformations
\[
\{ \Triangle \ITriangle, \Triangle \MTriangle \} \xrightarrow{U} \{ \UTileTriangle\MTriangle, \UTileTriangle\ITriangle  \}.
\]
As before, using the expressions given in previous sections for $U$ and for the Bell states, we obtain for the $U$-decorated entangled states the following expressions:
\begin{equation}
\begin{aligned}
    \ket{\UTileTriangle \Triangle} &= \frac{1}{2\sqrt{2}}\Bigl(\ket{\Circle \Circle} + \sqrt{3} \ket{\Circle\CIRCLE} + \sqrt{3} \ket{\CIRCLE\Circle} - \ket{\CIRCLE\CIRCLE}\Bigr) \\
    \ket{\UTileTriangle \ITriangle} &= \frac{1}{2\sqrt{2}}\Bigl(\sqrt{3}\ket{\Circle \Circle} +  \ket{\Circle\CIRCLE} -\ket{\CIRCLE\Circle} + \sqrt{3}\ket{\CIRCLE\CIRCLE}\Bigr) \\
    \ket{\UTileTriangle \MTriangle} &=\frac{1}{2\sqrt{2}}\Bigl(\ket{\Circle \Circle} + \sqrt{3} \ket{\Circle\CIRCLE} + \sqrt{3} \ket{\CIRCLE\Circle} + \ket{\CIRCLE\CIRCLE}\Bigr) \\
    \ket{\HTileTriangle \IMTriangle} &= \frac{1}{2\sqrt{2}}\Bigl(\sqrt{3}\ket{\Circle \Circle} +  \ket{\Circle\CIRCLE} +\ket{\CIRCLE\Circle} + \sqrt{3}\ket{\CIRCLE\CIRCLE}\Bigr).
\end{aligned}
\end{equation}
Using Born Rule one can easily verify the measurement probabilities given in Fig.~\ref{fig:ParallelepipedoU}.\footnote{In Fig.~\ref{fig:ParallelepipedoU} these probabilities have been approximated by excess and defect to the second decimal place, so that $\frac{1}{8} = 0,125 \simeq \frac{12}{100}$ and $\frac{3}{8} = 0,375 \simeq \frac{38}{100}$.} Notice that, as one could expect, $U$-decorated entangled states are not \emph{flat}, since the probabilities for the four possible outcomes are not all equal. For example, if the state is   $\UTileTriangle \IMTriangle$, one has a probability of $\frac{3}{4}$ of observing correlated results and a probability of $\frac{1}{4}$ of observing anti-correlated results; and the opposite happens for the state $\UTileTriangle \MTriangle$. However, if one looks only at the result for a single sub-system, it is easy to check that for all $U$-decorated entangled states the outcome is still completely random, as it is for $H$-decorated entangled states.

\subsection{Mixed States}
In many realistic scenarios it can happen that the state of a system is not completely known. For example, the preparation procedure may be affected by some experimental uncertainty due to interactions with the environment; or it may happen that one makes a measurement but does not look at the result, so that the post-measurement state is definitely one among the possible outcomes, but we don't know which one. In all such cases, the system is described by a \emph{mixed} state, also called sometimes \emph{classical mixture} of quantum states. In QM, using the formalism of density operators, pure states are represented by rank 1 projectors on the Hilbert space of the system and mixed states are represented by convex combinations of pure states. It is easy to see that, if the state is mixed, by applying a unitary operation the resulting state will still be mixed. Moreover, whenever one performs a measurement on a mixed state, in addition to the essentially "quantum" probabilities arising from superposition - which do not appear if one performs the measurement on an eigenstate of the observable to be measured - one must take into account also the "classical" probabilities - those arising from the uncertainty over the actual state of the system. The two probabilities come from different sources and are independent from each other, so they factorize for each state appearing in the mixture. As we will see, these properties of mixed states under operations and measurements are sufficient to extend QTris' rules and include mixed states into the game.

\begin{figure*} 
\centering\includegraphics[width=.8\textwidth]{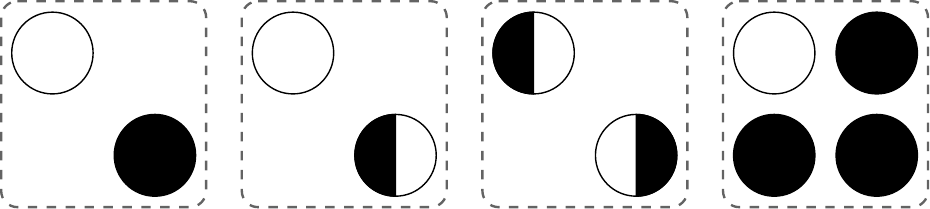}
    \caption{
    \textbf{Mixed states.} Four examples of mixed states. Both  $\protect \MixedOne$ and $\protect \MixedThree$ are completely mixed states and do not change under any card. They also give the same probabilities (1/2,1/2). The state $\protect \MixedTwo$ is not completely mixed and gives probabilities $(3/4, 1/4)$. It has limited ways of transforming with the cards. The state $\protect \MixedFour$ is mixed with probabilities $(1/4,3/4)$.}
    \label{fig:MixedStates}
\end{figure*}

First, we must define a notation for mixed states (see Fig.~\ref{fig:MixedStates}). Consider, for example, a cell which has probability $1/2$ to be in the state $\Circle$ and a probability $1/2$ to be in the state $\CIRCLE$. Such a mixed state will be represented as $\MixedOne$, that is, by putting one tile $\Circle$ and one tile $\CIRCLE$ within the cell. Another example: if a cell has probability $1/4$ of being in the state $\Circle$ and probability $3/4$ of being in the state $\CIRCLE$, it will be in the mixed state $\MixedFour$. Of course one can mix color tiles and orientation tiles: for example, if a cell has probability $1/2$ of being in the state $\Circle$ and probability $1/2$ of being in the state $\LEFTcircle$, it will be represented as $\MixedTwo$.\footnote{Of course this notation is limited to mixtures whose proportions are expressed by whole numbers; in the following, we will limit ourselves to such cases.} Notice that the arrangement of the tiles within a single cell in a mixed state is completely irrelevant.

The behavior of mixed states under the action of cards is the following: \emph{every time a card is used on a mixed state, it acts on all the tiles in the cell}. Thus, for example, if we play the card $X$ on a cell in the mixed state $\MixedFour$, we have the  transformation $
\MixedFour \xrightarrow{X} \XMixedFour$. 

As for measurement probabilities, as remarked above, we must take into account the probabilities arising from two independent sources: the first, coming from quantum superposition, is a feature of both pure and mixed states, while the second, coming from classical uncertainty over the actual state of the system, only appears if the state is mixed. In this latter case, the probability of obtaining a result $x$ under measurement is given by the weighted mean of the corresponding $x$ probabilities for the pure states appearing in the mixture, the weights being given by the probability distribution characterizing the mixture. For example, if the state is $\MixedTwo$, we have a probability $1/2$ for the state $\Circle$ and a probability $1/2$ for the state $\LEFTcircle$, so the weights are $1/2$ and $1/2$; then, each possibility contributes with its intrinsic probability of giving the result $\Circle$. In this case, as we know, the state $\Circle$ has probability $1$ to give the result $\Circle$, while the state $\LEFTcircle$ has probability $1/2$. Thus, in symbols:
\[ 
prob(\Circle | \MixedTwo) = \frac{1}{2}prob(\Circle|\Circle) + \frac{1}{2}prob(\Circle|\LEFTcircle) = \frac{1}{2}\cdot 1 + \frac{1}{2}\cdot \frac{1}{2} = \frac{3}{4}.
\]

As we mentioned above, if after a measurement one does not look at the result, one can only assign some probabilities to the possible post-measurement states. Thus, in this situation, the system after measurement will be described by a mixed state. We can use with circumstance to add a rule to QTris and make mixed states part of the game mechanics. The rule is the following: \emph{after every turn, one random cell is measured, but the result remains unknown}. The cell to be measured is chosen by rolling a d10: with an outcome between $1$ and $9$, the tile on the corresponding cell gets measured; with an outcome $0$, nothing happens. Now, as we already know, if the state of the measured cell is either $\Circle$ or $\CIRCLE$, it remains the same and the state is still pure as before; but if the tile is in a color superposition, the post-measurement state will be a mixture of color states. For example, if the state is either $\LEFTcircle$ or $\RIGHTcircle$, upon measurement we obtain one $\Circle$ or $\CIRCLE$ with probability \textonehalf; thus, we must place on that cell {\em both} color tiles $\Circle$ and $ \CIRCLE$ and the post-measurement state will be the completely mixed state. Similarly, if on the measured cell there is a decorated state like $\UTileCIRCLE$, we must put $3$ $\Circle$ and $1$ $\CIRCLE$ to make the mixture $\XMixedFour$, which  reflects the measurement probabilities of the initial state. 

Through the introduction of mixed state QTris game mechanics offers a clear, unambiguous and operational way to distinguish the properties of a superposition state from a classical mixture, one of the most delicate matters in teaching QM \cite{Passante}. In fact, it is easy to see that, for example, the behavior under measurement of the pure state $\LEFTcircle$ is indistinguishable from that of the mixed state $\MixedOne$. Therefore, how are we to distinguish them? The answer, as usual, is in  how they transform under unitary operators. If we play the $H$ card on $\LEFTcircle$, we obtain $\Circle$ and, therefore, we are certain of getting $\Circle$ upon measurement. Similarly, by playing $Z$ first and then $H$ we obtain $\CIRCLE$ with certainty. So, no matter what color one plays for, using the right cards there is a way to transform $\LEFTcircle$ into any other tile, thus getting a strategic advantage. On the other hand, if we play the $H$ card on the mixed state $\MixedOne$, we have $\Circle \xrightarrow{H} \LEFTcircle$ on the first tile, and $\CIRCLE \xrightarrow{H} \RIGHTcircle$ on the second tile, so that the new state is $\MixedThree$. Now, what are the measurement probabilities for this mixed state? Using the rule given above, $\LEFTcircle$ comes with probability \textonehalf\ and returns probability \textonehalf\ for both $\Circle$ and $\CIRCLE$, and the same holds for $\RIGHTcircle$. Thus we have:
\[ 
prob(\Circle | \MixedOne) = \frac{1}{2}prob(\Circle|\LEFTcircle) + \frac{1}{2}prob(\Circle|\RIGHTcircle) = \frac{1}{2}\cdot \frac{1}{2} + \frac{1}{2}\cdot \frac{1}{2} = \frac{1}{2}.
\]
So, all in all, applying $H$ we get a probability \textonehalf\ for both $\Circle$ and $\CIRCLE$, exactly as in the initial state. We then see that the card $H$, which has a fundamental function for the state $\LEFTcircle$, it is completely useless on the mixed state $\MixedOne$. In fact, it is easy to verify that for \emph{completely mixed states}, that is, mixtures of orthogonal states with all weights equal, there is no way to move away from complete randomness using unitary operations. In other terms, in contrast with superposition states, completely mixed states are \emph{useless} from the strategic point of view. Notice, however, that this is not the case if the state is mixed but not completely.    Consider, for example, the state $\MixedFour$, and what happens when one plays the card $X$ on it. The initial state $\MixedFour$ has probabilities $(1/4, 3/4)$. By using $X$, all the three $\CIRCLE$ flip as $\CIRCLE \xrightarrow{X} \Circle$, while the one $\Circle$ flips as $\Circle \xrightarrow{X} \CIRCLE$, thus resulting in the mixed state $\XMixedFour$ with three $\Circle$ and one $\CIRCLE$, with probabilities $(3/4, 1/4)$. Thus, not completely mixed states can be partially manipulated, but \emph{never} in such a way to get \emph{certain} results, as can be \emph{always} done with pure quantum states.

\section{Preliminary Empirical Findings}
In recent years QTris has been adopted by the National Quantum Science and Technology Institute (NQSTI) and used as a pedagogical tool to teach and illustrate the core concepts of QM by many NQSTI-affiliated researchers around Italy. Overall, since 2022 QTris has been played by hundreds of students, teachers and general audiences in a variety of formal, non formal and informal settings \cite{Bondani2026}. Among other things, NQSTI organized two on-line teacher training courses on QM based on QTris in 2023-2024 and incorporated the game in the curriculum of \emph{Quantum Science and Technology for High Schools}, a Summer School for in-service teachers held in Volterra in July 2025. Moreover, between winter 2024 and spring 2026 some of the authors conducted a pilot study for a QTris-based full course on QM involving three high-school classes in Naples; the design, implementation and results of this experimentation will be presented in depth in a future work.

Here we report on a stand-alone educational activity organized by Liceo "Galileo Galilei" in Siena in spring 2026. The initiative, promoted by one of the teachers who participated to  \emph{Quantum Science and Technology for High Schools}, involved approximately 150 students attending the fourth and fifth years of upper secondary school (16-18 years old) with no previous knowledge of QM. The program of the event included two complementary parts: in the first part a short seminar (about 60 minutes) introduced the basic principles of QM within QTris framework, emphasizing the concepts of state preparation, incompatible observables, measurements and the role of unitary operators; in the second part (about 60 minutes) students played the basic version of QTris. 
At the end of the activity, students completed a multiple-choice questionnaire  designed to evaluate their immediate understanding of the concepts presented during the seminar and assess the effectiveness of QTris as a tool to convey an immediate, basic understanding of these concepts, together with an operational ability to analyze simple game sequences. Participation was voluntary and all answers were analyzed anonymously.

\begin{table}[htbp]
\centering
\footnotesize
\renewcommand{\arraystretch}{1.2} 
\begin{tabularx}{\textwidth}{@{} c X X @{}}
\toprule
\textbf{Item} & \textbf{Question} & \textbf{Possible Answers} \\
\midrule
Q1 & What are the phases of a quantum experiment? & \begin{tabenum}
    \item \textbf{Preparation - Operations - Measurement}
    \item Design - Observation - Counting
    \item Preparation - Operations - Analysis
    \item Preparation - Measurement - Analysis
\end{tabenum} \\
\midrule

Q2 & The result of a single measurement is probabilistic & 
\begin{tabenum}
    \item \textbf{True}
    \item False
\end{tabenum} \\
\midrule

Q3 & The result of a single measurement is always predictable with certainty & 
\begin{tabenum}
    \item True
    \item \textbf{False}
\end{tabenum} \\
\midrule

Q4 & The result of a single measurement is always undetermined & 
\begin{tabenum}
    \item True
    \item \textbf{False}
\end{tabenum} \\
\midrule

Q5 & The result of a single measurement is in some cases predictable with certainty & 
\begin{tabenum}
    \item \textbf{True}
    \item False
\end{tabenum} \\
\midrule

Q6 & $C$ and $O$ are two maximally incompatible physical properties. If, for a given physical system, the result of a measurement of $C$ is known with certainty, the result of a measurement of $O$ will be... & 
\begin{tabenum}
    \item Equal to that of $C$
    \item Opposite to that of $C$
    \item Known with certainty
    \item \textbf{Maximally undetermined}
\end{tabenum} \\
\midrule

Q7 & In QM the role of unitary operators is... & 
\begin{tabenum}
    \item To describe a measurement process
    \item \textbf{To transform a quantum state}
    \item To give a probability
    \item To select a measurement outcome
\end{tabenum} \\
\midrule

Q8 & Consider the following sequence of QTris' operations: 
\begin{tabenum}
    \item Prepare a square with a white tile
    \item Apply the operation $X$
    \item Measure the color of the tile
\end{tabenum} & 
\begin{tabenum}
    \item White or Black, with probabilities $50\%$ and $50\%$
    \item White or Black, with probabilities $25\%$ and $75\%$
    \item Left or Right, with probabilities $50\%$ and $50\%$
    \item \textbf{Black with certainty}
\end{tabenum} \\
\midrule

Q9 & Consider the following sequence of QTris' operations: 
\begin{tabenum}
    \item Prepare a square with a black tile
    \item Apply the operation $X$
    \item Apply the operation $H$
    \item Measure the color of the tile
\end{tabenum}& 
\begin{tabenum}
    \item \textbf{White or Black, with probabilities $50\%$ and $50\%$}
    \item White or Black, with probabilities $25\%$ and $75\%$
    \item Left or Right, with probabilities $50\%$ and $50\%$
    \item White with certainty
\end{tabenum} \\
\midrule

Q10 & Consider the following sequence of QTris' operations: 
\begin{tabenum}
    \item Prepare a square with a right tile
    \item Apply the operation $Y$
    \item Apply the operation $H$
    \item Measure the color of the tile
\end{tabenum}& 
\begin{tabenum}
    \item \textbf{White with certainty}
    \item White or Black, with probabilities $50\%$ and $50\%$
    \item Left or Right, with probabilities $50\%$ and $50\%$
    \item Black with certainty
\end{tabenum} \\
\midrule

Q11 & Consider the following sequence of QTris' operations: 
\begin{tabenum}
    \item Prepare a square with a left tile
    \item Apply the operation $H$
    \item Measure the orientation of the tile
\end{tabenum}& 
\begin{tabenum}
    \item Right with certainty
    \item Black with certainty
    \item White or Black, with probabilities $50\%$ and $50\%$
    \item \textbf{Left or Right, with probabilities $50\%$ and $50\%$}
\end{tabenum} \\
\midrule

Q12 & Consider the following sequence of QTris' operations: 
\begin{tabenum}
    \item Prepare a square with a right tile
    \item Apply the operation $Z$
    \item Measure the orientation of the tile
\end{tabenum}& 
\begin{tabenum}
    \item Right with certainty
    \item Left or Right, with probabilities $50\%$ and $50\%$
    \item \textbf{Left with certainty}
    \item Black with certainty
\end{tabenum} \\

\bottomrule
\end{tabularx}

\caption{The questionnaire given at the end of a QTris-based activity at Liceo "Galileo Galilei". In bold the correct answers.}
\label{tab:questionnaire}
\end{table}

The questionnaire consisted of twelve multiple choice questions organized into two complementary domains. The first seven items focused on foundational concepts introduced during the seminar, including the structure of a quantum experiment (Q1), the probabilistic nature of quantum measurements (Q2-5), incompatible properties (Q6) and the role of unitary operators (Q7). The remaining five questions (Q8-Q12) required students to apply these concepts and QTris' rules to solve simple game problems requiring to predict the outcome of short sequences of operations and measurements.

\begin{figure}
    \centering
\includegraphics[width=\textwidth]{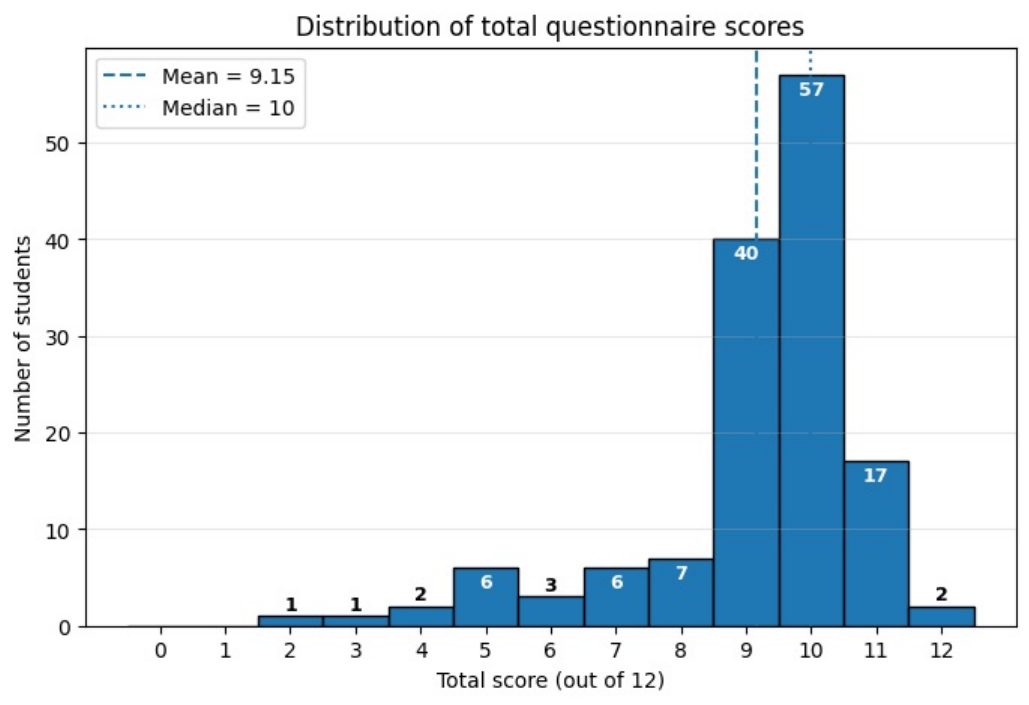}
    \caption{Score distribution over participants.}
    \label{fig:score_distribution}
\end{figure} 

\begin{figure}
    \centering
\includegraphics[width=\textwidth]{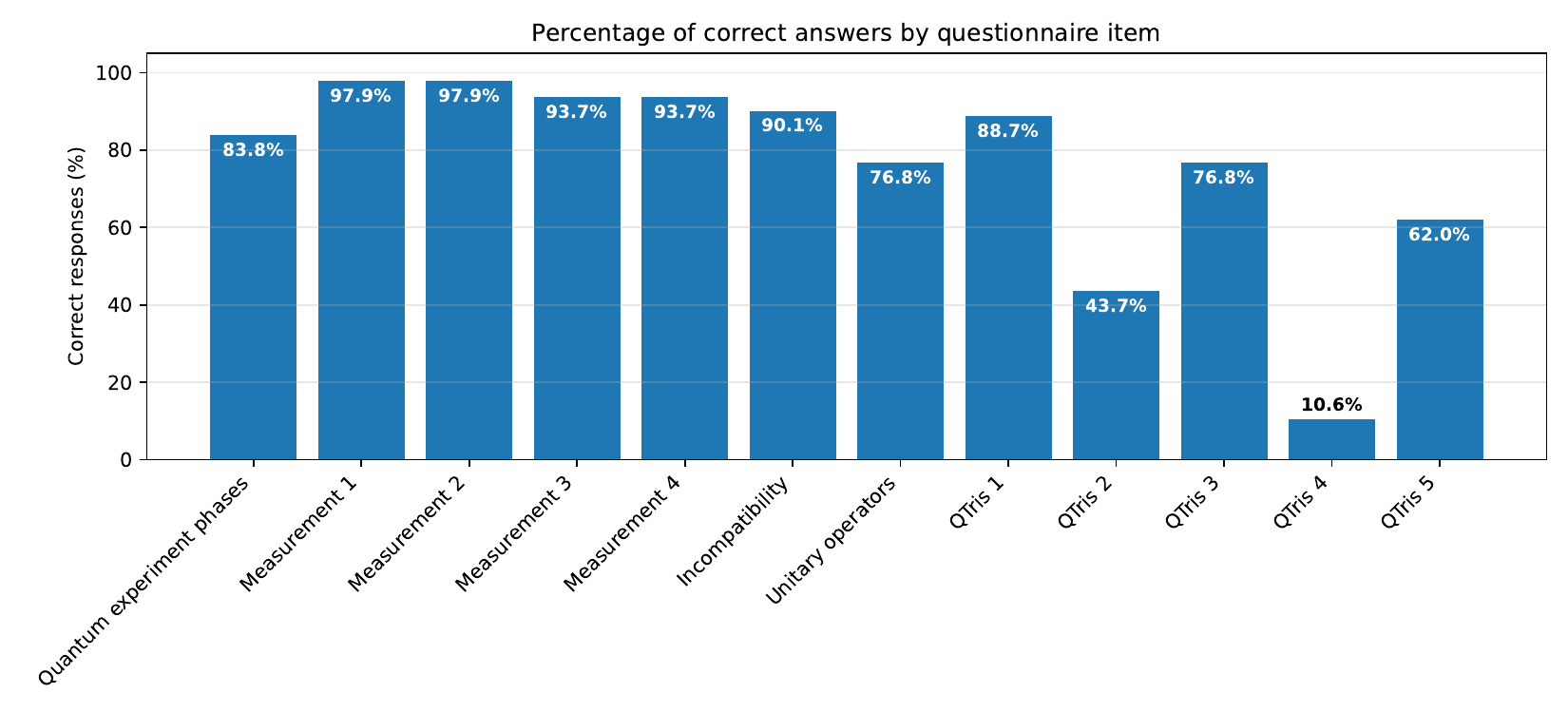}
    \caption{Percentage of correct answers for each question.}
    \label{fig:accuracy_by_question}
\end{figure} 

\begin{figure}
    \centering
\includegraphics[width=0.75\textwidth]{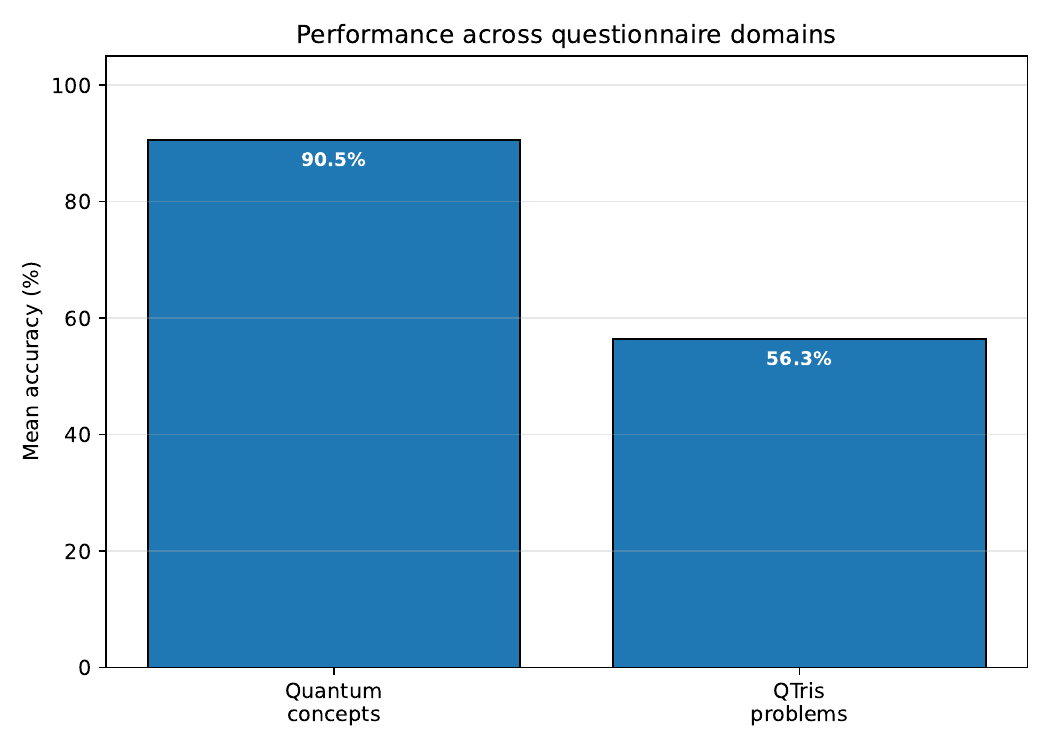}
    \caption{Percentage of correct answers to conceptual and operational questions.}
    \label{fig:concept_vs_operational}
\end{figure} 

\begin{figure}
    \centering
\includegraphics[width=\textwidth]{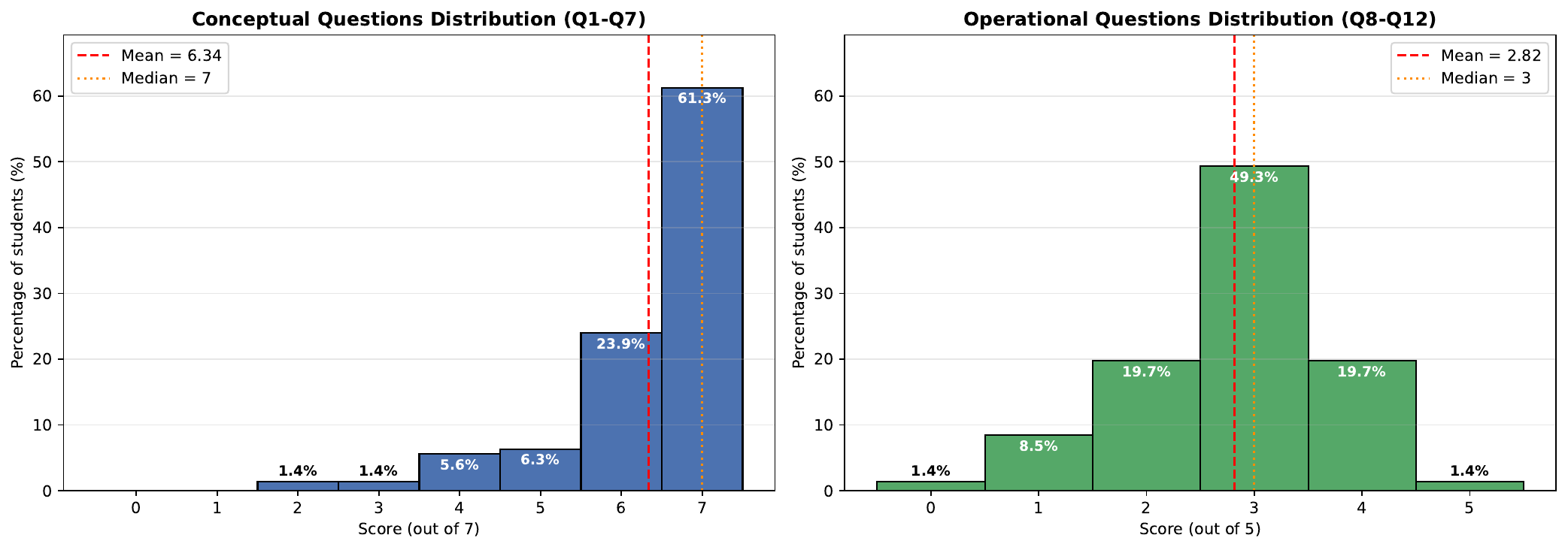}
    \caption{Distribution of scores for conceptual and operational questions.}
\label{fig:concept_vs_operational_detail}
\end{figure} 

A total of $142$ valid questionnaires were collected and included in the analysis. The distribution of total scores is shown in Figure~\ref{fig:score_distribution}. Most students achieved high overall performances, with a mean score of $9.15$ out of $12$, a median score of $10$ and a standard deviation of $1.74$. These results indicate a generally high level of immediate understanding across participants, together with sufficient variability to discriminate different levels of performance. The absence of strong ceiling or floor effects indicates that the questionnaire provides a reasonable discrimination between different levels of immediate understanding.

\begin{figure}
    \centering
\includegraphics[width=0.75\textwidth]{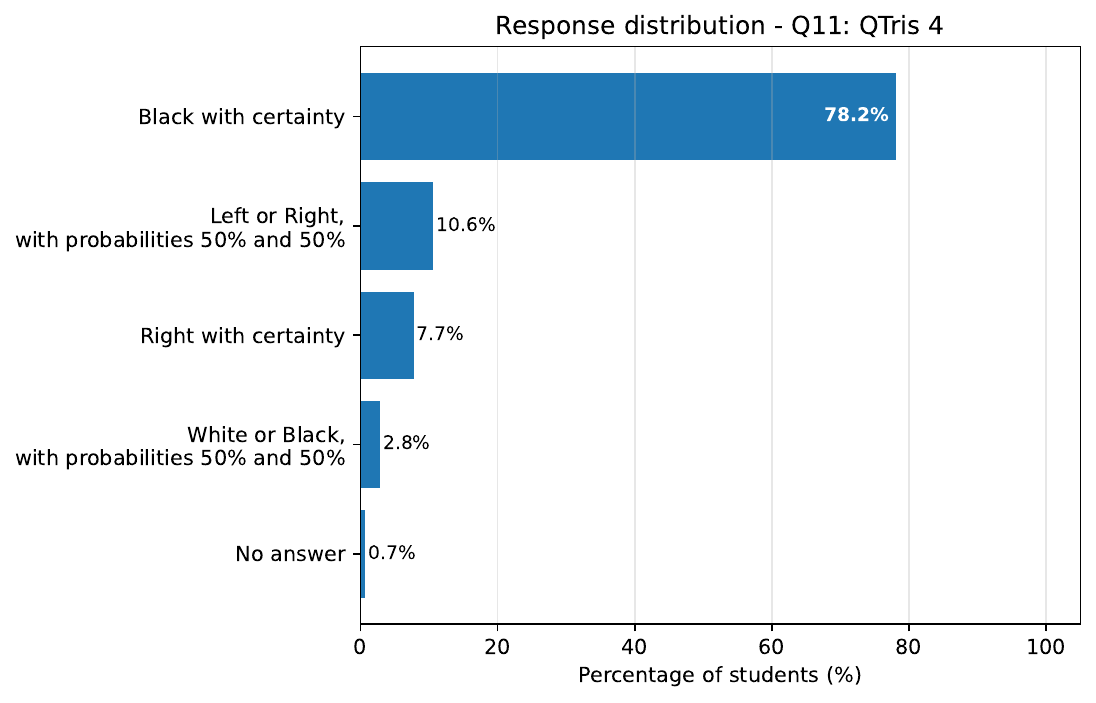}
    \caption{Distribution of answers for item Q11.}
\label{fig:Q11}
\end{figure} 

\begin{figure}
    \centering
\includegraphics[width=0.75\textwidth]{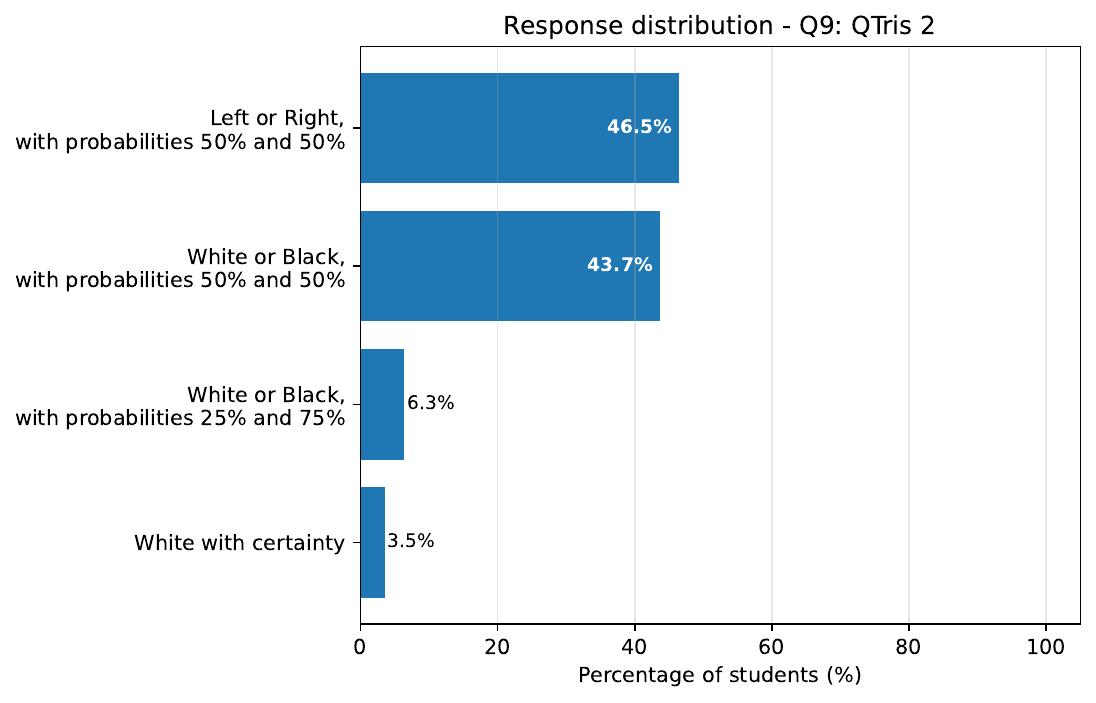}
    \caption{Distribution of answers for item Q9.}
\label{fig:Q9}
\end{figure}

Figure~\ref{fig:accuracy_by_question} shows the percentage of correct answers for each questionnaire item. The overall results of the questionnaire provide encouraging indication that QTris effectively supported an immediate acquisition of some of the key concepts highlighted in Section 3. In particular, students demonstrated a strong understanding of the structure of quantum experiments (Q1, 83.8\%), probabilistic measurements (Q2-Q5, average 95.8\%), meaning of \emph{incompatibility} (Q6, 90.1\%) and role of unitary transformations (Q7, 76.8\%).
On the other hand, in QTris-based operational questions students generally demonstrated a good understanding of the basic game mechanics, but more complex sequences involving multiple operations and a final measurement proved more challenging. As shown in Figure~\ref{fig:concept_vs_operational}, there is a significant gap in the results for the conceptual and operational parts of the questionnaire (90.5\% vs 56.3\%). The separate distributions of scores for the two parts are shown in Figure~\ref{fig:concept_vs_operational_detail}. While more than 60\% of students answered correctly to all $7$ conceptual questions, and more than 80\% to at least $6$ questions, for the operational questions we have a mean score of $2.83$ and a median of $3$ out of $5$.  This could be expected, since the correct answers to the conceptual questions were given during the seminar, while translating quantum concepts into operational, problem-solving abilities may require a longer practice with the rules of the game. The lower trend is particularly evident for item Q11, to which only the 10.6\% of participants answered correctly. This question required students to predict the result of an Orientation measurement on a Color tile, i.e., to virtually perform a change of basis. Figure~\ref{fig:Q11} shows the distribution of answers to this item. Most students (78.2\%) answered "Black with certainty", which is the correct answer if the final measurement was a Color rather than an Orientation measurement. It is possible that many students were confused by the operation of measuring Orientation on a Color tile, something which, indeed, goes beyond the immediate understanding of the game mechanics, where only Color measurement are made. However, the main difficulty does not seem to be in Orientation measurements \emph{per se}, in light of the better performance observed for item Q12 (62.0\%). This problem still involves an Orientation measurement, but it is performed on an Orientation tile and thus gives a certain outcome. These results suggests that measurement in a different basis represented the most cognitively demanding aspect of the activity, something probably requiring more time and practice than available in a stand-alone event. In this regard, it is noteworthy that also item Q9 shows a mean score lower than the average ($43.7\%$). Looking at the distribution of answers to this item (Figure~\ref{fig:Q9}) it seems that most students (90.2\%) realized that the result was maximally \emph{random} - as suggested by QTris' (Color) measurement rules - but about half of them answered "Right or Left, with probabilities 50\% and 50\%"; i.e., they associated the correct probabilities to the wrong outcomes. Also here, the main confusion seems to arise in connection with measurements performed on states which are not eigenstates of the measured observable. 

The overall results of the questionnaire suggest that QTris is capable of introducing abstract quantum concepts through an intuitive game-based approach and of promoting for a significant part of students some procedural reasoning analogous to that required when manipulating quantum states in elementary quantum-information protocols. Although the present preliminary analysis evaluates only the immediate impact of a stand-alone educational activity, we believe that the overall performance obtained by students supports the use of QTris as an effective framework to introduce the basic concepts of QM.

\section{Conclusions and Future Work}
In this paper we introduced the new version of QTris, a board game designed to teach and learn QM within the framework of Quantum Information Theory and Quantum Computation.  Despite our main focus here was on high school teaching, we believe that QTris may become a useful tool for everyone engaged in the difficult task to teach and explain QM in different contexts and at different levels. From outreach events to workforce training, it is a flexible platform that can be adapted to a variety of needs.

The key idea behind the game, which is a variant of the traditional \emph{Tic-Tac-Toe}, is that the board represents a system of qubits that can be in different states. Players use cards corresponding to unitary operators to transform states in a deterministic way and, at the end of the game,  a (probabilistic) measurement is made to assign points to players. So, in short, the aim of the players is to use the cards at their disposal to maximize their probability to score more points than their opponent. 

Thanks to the structural analogy described in Section 5, which links QTris' game elements and QM mathematical framework for a two-state system, by design every QTris' game sequence simulates a possible quantum process made of \emph{preparation}, \emph{unitary transformations} and \emph{measurement}. Thus, QTris provides an operational framework to familiarize with the postulates of QM and their consequences, when they are applied to a system of qubits. In particular, the game mechanics is designed in such a way to highlight some key concepts of QM, namely the existence of incompatible properties and the mixed character of the theory, which includes both probabilistic and deterministic aspects. Moreover, by the same structural analogy, we saw how every game sequence can be easily translated into algebraic expressions using Dirac notation, so that one can easily frame QTris' game problems that are actually typical problems in QM. Thus, QTris becomes a pedagogical platform to practice with linear algebra, probability theory and the postulates of QM, allowing students to \emph{use} QM to solve problems, just as they are used to do in classical physics. At the same time, and from a higher perspective, QTris can be analyzed within the framework of QGT, since it satisfies all the essential criteria of quantum games. 

As we saw in Section 6, one of the most valuable features of QTris is that it can be easily expanded by adding new states and operations, so to incorporate new phenomenology into the game mechanics. Here we described how to introduce entanglement, partial incompatibility and mixed states in the game, and additional extensions will be described in future works. In this regard, let us mention another important theoretical element of QM which, at present, is not included in QTris' game mechanics, that the infinitesimal generator of the time evolution for quantum states is the Hamiltonian operator associated to the energy of the system. An additional extension of the game including this important element is currently under development. A further upgrade to QTris' pedagogical potential will come from \emph{QTris App}, a digital version of the game currently under development. In particular, the possibility to perform a great number of repeated measurements will allow to explore the important concepts of expectation value and variance of an observable. 

Finally, in Section 7, we reported on a QTris-based educational activity that involved about 150 high-school students. The results of our analysis are only a preliminary step, but they provide encouraging indications that QTris can be effectively used as a tool to convey an immediate understanding of some key concepts of QM, together with a good operational ability to apply these concepts within QTris' framework. Also these aspects will be analyzed in greater detail in a future work, where we will report on the design, implementation and results of three full QTris-based introductions to QM.

\section*{Declarations}

\subsection*{Ethics Approval}
Not applicable.

\subsection*{Competing interests}
The authors declare no competing interests.

\subsection*{Dual publication}
The results/data/figures in this manuscript have not been published elsewhere, nor are they under consideration by another publisher.

\subsection*{Authorship}
I confirm the corresponding author has read the journal policies and submit this manuscript in accordance with those policies.

\subsection*{Third party material}
All the material is owned by the authors and no permissions are required.

\subsection*{Availability of data and materials}
The data that support the findings of this study are available upon request and with the permission University of Naples.

\subsection*{Funding}
No funding to be declared.

\subsection*{Authors' contributions} 
A.H. owns the copyright of QTris game (protected under SIAE registry No. 2024/00042); A.A., A.H. and M.B. wrote the main manuscript text; A.H., M.V., I.D.S. and M.N. coauthored QTris’ rulebook, M.V. and I.D.S. prepared figures; M.N. collected and analyzed data; all authors reviewed the manuscript.

\subsection*{Acknowledgements}
The Authors acknowledge the collaboration of Professor Angela Santoriello and students of Liceo Scientifico  "Galileo Galilei" (Siena) who made possible the activity  described in Section 7.


\begin{thebibliography}{00}

\bibitem{Krijt2017} Krijtenburg-Lewerissa K, Pol HJ, Brinkman A, van Joolingen WR. Insights into teaching quantum mechanics in secondary and lower undergraduate education. Phys Rev Phys Educ Res. 2017;13(1):010109.

\bibitem{Merzel2024a} Merzel A, et al. The core of secondary level quantum education: a multi-stakeholder perspective. EPJ Quantum Technol. 2024;11:27.

\bibitem{flagship} Quantum Flagship. https://qt.eu/.

\bibitem{Stadermann2019} Stadermann HKE, van den Berg E, Goedhart MJ. Analysis of secondary school quantum physics curricula of 15 different countries: Different perspectives on a challenging topic. Phys Rev Phys Educ Res. 2019;15:010130.

\bibitem{Verbraeken2024} Verbraeken L, et al. Towards a Quantum Technology PCK for Teachers. J Phys Conf Ser. 2024;2750:012045.

\bibitem{Chiofalo2022} Chiofalo ML, Foti C, Michelini M, Santi L, Stefanel A. Games for teaching/learning quantum mechanics: a pilot study with high-school students. Educ Sci. 2022;12:446.

\bibitem{Watrous} Watrous J. The Theory of Quantum Information. Cambridge: Cambridge University Press; 2018.

\bibitem{NielsenChuang} Nielsen MA, Chuang IL. Quantum Computation and Quantum Information: 10th Anniversary Edition. Cambridge: Cambridge University Press; 2010.

\bibitem{Bondani2024} Bondani M, Caprara S, Chiarello F, Dabbicco M, Hamma A, Malgieri M, et al. Quantum Technologies 2024. Proc SPIE. 2024;12993:129930.

\bibitem{Bondani2026} Bondani M, et al. QTris, a new game for teaching quantum physics. J Phys Conf Ser. 2026;3203:012035.

\bibitem{qtris} NQSTI. QTris - Il gioco della meccanica quantistica. https://nqsti.it/news/qtris-il-gioco-della-meccanica-quantistica.


\bibitem{Giliberti2024} Giliberti M, Lovisetti L. Old Quantum Theory and Early Quantum Mechanics. Cham: Springer; 2024.

\bibitem{Singh2008} Singh C. Student understanding of quantum mechanics at the beginning of graduate instruction. Am J Phys. 2008;76:277-287.

\bibitem{Fischler} Fischler H, Lichtfeldt M. Modern physics and students' conceptions. Int J Sci Educ. 1992;14(2):181-190.

\bibitem{DuncanTrilogy} Duncan A. Von Neumann's 1927 trilogy on the foundations of quantum mechanics: Annotated translations. arXiv:2406.02149. 2025.

\bibitem{VonNeumann} von Neumann J. Mathematical Foundations of Quantum Mechanics: New Edition. Princeton: Princeton University Press; 2018.

\bibitem{DuncanJanssen1} Duncan A, Janssen M. Constructing Quantum Mechanics, Volume 1: The Scaffold: 1900-1923. Oxford: Oxford University Press; 2019.

\bibitem{DuncanJanssen2} Duncan A, Janssen M. Constructing Quantum Mechanics, Volume 2: The Arch: 1923-1927. Oxford: Oxford University Press; 2023.

\bibitem{OxfordPrincipia} Buchwald JZ, Fox R. Newton's Principia. In: Smeenk C, Schliesser E, editors. The Oxford Handbook of the History of Physics. Oxford: Oxford University Press; 2013. p. 109-165.

\bibitem{Guicciardini} Guicciardini N. Reading the Principia: The Debate on Newton's Mathematical Methods for Natural Philosophy from 1687 to 1736. Cambridge: Cambridge University Press; 1999.

\bibitem{Nauenberg} Nauenberg M. The reception of Newton's Principia. arXiv:1503.06861. 2015.

\bibitem{Hestenes} Hestenes D, Wells M, Swackhamer G. Force concept inventory. Phys Teach. 1992;30(3):141-158.

\bibitem{DeGandt} De Gandt F. Force and Geometry in Newton's Principia. Princeton: Princeton University Press; 2016.

\bibitem{QTEdu} European Quantum Readiness Center (QTEdu). https://qtedu.eu/.

\bibitem{competence} Greinert F, M\"uller R. Competence framework for quantum technologies: methodology and version history. Brussels: European Commission; 2021.

\bibitem{Michelini2000} Michelini M, Ragazzon R, Santi L, Stefanel A. Proposal for quantum physics in secondary school. Phys Educ. 2000;35(6):406.

\bibitem{McIntyre2022} McIntyre DH. Quantum Mechanics: A Paradigms Approach. Cambridge: Cambridge University Press; 2022.


\bibitem{Merzel2024b} Merzel A, Weissman EY, Katz N, Galili I. Mathematical structures in quantum physics education for high school students: Unveiling the power of Dirac notation for conceptual and problem-solving proficiency. Phys Rev Phys Educ Res. 2024;20:020134.


\bibitem{Toth} T\'oth K, et al. Investigating the effect of two-state approaches on students' understanding of quantum measurement: A quasiexperimental field study. Phys Rev Phys Educ Res. 2025;21(2):020142.

\bibitem{Albert} Albert C, Förster M, Pospiech G. Developing a quantum physics curriculum for lower secondary education: insights into the design of a Spin First teaching concept and first empirical findings. EPJ Quantum Technol. 2025;12:132.

\bibitem{Bitzenbauer2024} Bitzenbauer P, et al. Design and evaluation of a questionnaire to assess learners' understanding of quantum measurement in different two-state contexts: The context matters. Phys Rev Phys Educ Res. 2024;20(2):020136. doi:10.1103/PhysRevPhysEducRes.20.020136.


\bibitem{Plass2015} Plass JL, Homer BD, Kinzer CK. Foundations of game-based learning. Educ Psychol. 2015;50(4):258-283.

\bibitem{Meyer1999} Meyer DA. Quantum strategies. Phys Rev Lett. 1999;82:1052.

\bibitem{Piispanen2025} Piispanen L, Pfaffhauser M, Wootton J, Togelius J, Kultima A. Defining quantum games. EPJ Quantum Technol. 2025;12:7.

\bibitem{Nawaz2004} Nawaz A, Toor AH. J Phys A Math Gen. 2004;37:4437.
\bibitem{Iqbal2002} Iqbal A, Toor AH. Phys Lett A. 2002;65:541.

\bibitem{D'Ariano2002} D'Ariano GM, Gill RD, Keyl M, Kuemmerer B, Maassen H, Werner RF. Quant Inf Comput. 2002;2:355.

\bibitem{qplaylearn} QPlayLearn. https://qplaylearn.com/.

\bibitem{goff2006} Goff A. Am J Phys. 2006;74:962.

\bibitem{Weingartner2023} Weingartner M, Weingartner T. Computers and Education Open. 2023;4:100125.

\bibitem{Kraus} Kraus K. States, Effects, and Operations: Fundamental Notions of Quantum Theory. Berlin: Springer; 1983.

\bibitem{StanfordUncertainty} Hilgevoord J, Uffink J. The uncertainty principle. In: Zalta EN, Nodelman U, editors. The Stanford Encyclopedia of Philosophy. Spring 2024 ed. 2024.

\bibitem{Eberhard} Eberhard PH. Bell's theorem and the different concepts of locality. Nuovo Cimento B. 1978;46:392-419. doi:10.1007/BF02728628.

\bibitem{Ghirardi} Ghirardi GC, Rimini A, Weber T. A general argument against superluminal transmission through the quantum mechanical measurement process. Lett Nuovo Cimento. 1980;27(10):293-298.

\bibitem{nonlocality} Brunner N, Cavalcanti D, Pironio S, Scarani V, Wehner S. Bell nonlocality. Rev Mod Phys. 2014;86:419.

\bibitem{Sciarrino} De Angelis T, Nagali E, Sciarrino F, De Martini F. Experimental test of the no-signaling theorem. Phys Rev Lett. 2007;99:193601.

\bibitem{Susskind} Susskind L, Friedman A. Quantum Mechanics: The Theoretical Minimum. New York: Basic Books; 2014.

\bibitem{Eisert1999} Eisert J, Wilkens M, Lewenstein M. Quantum games and quantum strategies. Phys Rev Lett. 1999;83(15):3077-3080. doi:10.1103/PhysRevLett.83.3077.

\bibitem{Marinatto2000} Marinatto L, Weber T. A quantum approach to static games of complete information. Phys Lett A. 2000;272(5):291-303. doi:10.1016/S0375-9601(00)00441-2.

\bibitem{BostanciWatrous2022} Bostanci J, Watrous J. Quantum game theory and the complexity of approximating quantum Nash equilibria. Quantum. 2022;6:882. doi:10.22331/q-2022-12-22-882.

\bibitem{Passante} Passante G, Emigh PJ, Shaffer PS. Student ability to distinguish between superposition states and mixed states in quantum mechanics. Phys Rev Spec Top Phys Educ Res. 2015;11:020135.

\end{thebibliography}
\end{document}